\documentclass[11pt]{article}
\usepackage[T1]{fontenc}
\usepackage{amsmath,amssymb,amsthm}
\usepackage[margin=1in]{geometry}
\usepackage{graphicx}
\usepackage{booktabs}
\usepackage[colorlinks=false,pdfborder={0 0 0}]{hyperref}

\newcommand{\dd}{\mathrm{d}}
\newcommand{\ee}{\mathrm{e}}
\newcommand{\ii}{\mathrm{i}}
\newcommand{\MP}{M_{\mathrm{P}}}
\renewcommand{\Box}{\square}

\title{Torsion, Frames and Quantum Bounces: Inflation and Ekpyrotic Contraction in Quantum Trajectory Cosmology with a Non-Minimally Coupled Scalar}
\author{
\begin{tabular}{c}
Rikpratik Sengupta$^{1}$ and Paulo Vargas Moniz$^{2}$\\[4pt]
\small $^{1}$Department of Physics, Indian Institute of Technology Kanpur, Kalyanpur 208016, India.\\
\small $^{2}$Departamento de F\'{\i}sica, Centro de Matem\'atica e Aplica\c{c}\~oes (CMA-UBI), Universidade da Beira Interior,\\
\small Rua Marqu\^{e}s d'\'Avila e Bolama, 6200-001 Covilh\~a, Portugal
\end{tabular}
}
\date{}

\begin{document}
\maketitle

\begin{abstract}
We study the de Broglie--Bohm quantum cosmology of a scalar field that is non-minimally coupled to a metric-compatible Riemann--Cartan geometry, and we use it to compare inflation with ekpyrotic contraction in one setting. The torsion trace is an auxiliary field and is eliminated algebraically before quantisation. The reduced two-dimensional minisuperspace metric depends on the coupling function $F(\phi)$ and on one further constant $\kappa$ that measures an independent quadratic torsion term, with $\kappa=0$ the Einstein--Cartan theory and $\kappa\to\infty$ the torsion-free theory. For every $F$, $\kappa$ and potential this metric is conformal to a flat metric, so that torsion acts only through the Einstein-frame kinetic function $K(\phi)$. The curvature of the minisuperspace metric is therefore not a diagnostic of torsion, and the Laplace--Beltrami Wheeler--DeWitt equation, its quantum potential and its Bohmian trajectories are covariant under the change of frame. For a massless scalar the Wheeler--DeWitt equation is solved exactly. In a two-packet state all 41 flux-weighted Bohmian trajectories that we integrated bounce in the Einstein frame. In the Jordan frame the bounce at the symmetric event does not depend on torsion and persists only for $\xi<\xi_c$, with $\xi_c$ given in closed form, and torsion changes the late-time expansion exponent from $1-\sqrt{6\xi/(1+6\xi)}$ to $1-\sqrt{6\xi}$. The bounce is produced by interference in the chosen state and not by torsion. For exponential potentials we give the classical scaling solution and the exact Wheeler--DeWitt solutions; in the sector that contains the classical ekpyrotic attractor no Bohmian trajectory bounces, and single-momentum states bounce only where the potential is negligible and always end in the crunch. For quadratic inflation the torsion-specific shift of the tensor-to-scalar ratio is below one per cent in the allowed range of $\xi$. A single ekpyrotic field gives blue spectra with $n_s=3+2/(\epsilon-1)$ and $r=16\epsilon$ whatever the torsion, which is confirmed numerically for a running $\epsilon$. In an entropic two-field version the tilt shifts at order $\sqrt\xi$ through the non-minimal coupling and at order $\xi^{3/2}$ through torsion. Tensor modes pass through the massless Bohmian bounce with transfer coefficients close to unity and stay blue.
\end{abstract}


\section{Introduction}
\label{sec:intro}

Quantum cosmology aims to describe the early universe at a stage where the classical Einstein equations cannot be trusted, and its simplest setting is the minisuperspace obtained by restricting the geometry and the matter to a few homogeneous degrees of freedom. The dynamics is then encoded in the Wheeler--DeWitt equation \cite{DeWitt1967}, supplemented by a choice of boundary condition for the wave function of the universe \cite{HartleHawking1983,Vilenkin1982}, and a general account of the formalism and its interpretational difficulties can be found in Refs.~\cite{Kiefer2007,Moniz2010}. Two difficulties are particularly relevant for what follows. The first is the ambiguity in the factor ordering of the Hamiltonian constraint and in the identification of a physical inner product, which affects the self-adjointness of the resulting operator and therefore the probabilistic interpretation \cite{HawkingPage1986,Halliwell1988,Almeida2015,Almeida2017}. The second is the absence of an external observer and of a preferred time, which is the motivation for the de Broglie--Bohm formulation \cite{Bohm1952}, in which the wave function guides a definite trajectory in configuration space without reference to measurement \cite{PintoNeto2013}. In this formulation the quantum potential supplies a repulsive contribution that can stop the collapse of the scale factor, so that the singularity of the classical solution may be replaced by a bounce, as was found for perfect fluids and for scalar fields \cite{deBarros1998,Alvarenga2002,Colistete2000}. Bouncing backgrounds of this kind have been supplemented with a theory of cosmological perturbations on the Bohmian trajectories, which leads to nearly scale invariant spectra in the matter bounce and shows how the quantum-to-classical transition of the perturbations can be described \cite{Peter2007,PeterPN2008,FalcianoPN2009,PNSS2012,PeterPNV2016}. Singularity avoidance has also been established in loop quantum cosmology \cite{Bojowald2001,Ashtekar2006,AshtekarSingh2011} and in relational quantisations \cite{Gielen2016}, and general reviews of bouncing cosmologies are available \cite{NovelloBergliaffa2008,BrandenbergerPeter2017}. These results have a common caveat, which is that whether a given quantisation resolves the singularity can depend on the choice of clock or of the variables in which the evolution is described \cite{GielenMP2020}, so that a claim of a bounce has to state in which variables it is made.

The same caveat arises for theories in which the gravitational action contains a non-minimal coupling between a scalar field and the curvature. The minisuperspace of such scalar--tensor theories has two degrees of freedom with a kinetic metric that contains a mixed term between the scale factor and the scalar field, and the quantum cosmology of these models was developed for the Brans--Dicke theory and for non-minimally coupled scalars by means of conformal transformations \cite{Fabris1999,Colistete2000}. Quantum cosmological scenarios of Brans--Dicke gravity have been compared in the Einstein and Jordan frames \cite{Almeida2018}, and the self-adjointness of the Brans--Dicke minisuperspace Hamiltonian has been studied \cite{Almeida2017}. The frame equivalence of the classical theory, which is understood in detail \cite{Faraoni1999}, need not carry over automatically to the quantum theory. The same Einstein-frame structure underlies the standard analysis of inflation with a non-minimal coupling \cite{Salopek1989,Futamase1989,Fakir1990,Bezrukov2008}.

Torsion offers a geometric route to modified singular behaviour. In Einstein--Cartan theory the connection is metric compatible but not symmetric, and its antisymmetric part is sourced by the spin of matter \cite{Kibble1961,Sciama1964,Hehl1976,Shapiro2002}. Trautman and Kopczy\'nski observed that the resulting spin--spin repulsion can replace the cosmological singularity by a bounce \cite{Trautman1973,Kopczynski1972}, and it was pointed out that this conclusion relies on the high symmetry of the model \cite{StewartHajicek1973}. The mechanism has been developed further for Dirac fermions coupled to torsion, where it leads to a non-singular big bounce and to an alternative to inflation \cite{Poplawski2010,Poplawski2012}. For a scalar field the situation is different, because the source of torsion is not spin but the gradient of the field multiplied by the derivative of the non-minimal coupling \cite{JhaLordSinha1988,JhaLordSinha1988b,Cid2018}. The torsion then has no kinetic term and can be eliminated algebraically, and the resulting metric theory is a scalar--tensor theory whose kinetic function depends on the torsion couplings \cite{Shaposhnikov2020,Langvik2021}. Homogeneous and isotropic torsion has only a vector and an axial component \cite{Tsamparlis1979}, and for a non-minimally coupled scalar the vector component is the one that is excited. The resulting Einstein--Cartan inflation is closely related to the Palatini formulation, in which the connection is independent of the metric and the inflationary observables depend on the coupling differently from the metric formulation \cite{Bauer2008,Shaposhnikov2021,Tenkanen2020}. In quantum cosmology, torsion-like structure has been studied in the teleparallel and symmetric teleparallel settings \cite{Cai2016,Dimakis2024fT,Dimakis2021fQ}, where the minisuperspace Hamiltonian is derived from a modified gravitational action, but the quantum cosmology of the Riemann--Cartan scalar--tensor theory with the torsion eliminated, and in particular its Bohmian interpretation, has not been worked out.

Contracting phases offer a second alternative to inflation. In the ekpyrotic and cyclic scenarios the universe contracts with an equation of state $w>1$ driven by a steep negative potential \cite{Khoury2001,SteinhardtTurok2002,Lehners2008}, a regime in which anisotropies and chaotic mixmaster behaviour are suppressed \cite{Erickson2004} and which is an attractor of the contracting phase \cite{Gratton2004}. A single scalar field in such a phase produces a strongly blue spectrum \cite{Lyth2002,Finelli2002}, and a nearly scale invariant spectrum requires the conversion of the perturbation of a second field, the entropic mechanism \cite{Lehners2007,Buchbinder2007,Creminelli2007,Koyama2007}. How inflation and ekpyrotic contraction compare when the same gravitational sector, including torsion and a non-minimal coupling, is used for both, and how the comparison extends to the quantum level, has not been examined.

The question addressed here is how torsion of a four-dimensional parent theory is represented once the minisuperspace Wheeler--DeWitt equation and its Bohmian interpretation are constructed, which parts of the resulting dynamics are due to torsion and which to the non-minimal coupling alone, and what this implies for inflation and for ekpyrotic contraction. The second part requires a control comparison, which is built into the model. The action contains the coupling function $F(\phi)$ and a constant $\kappa\ge0$ that multiplies an independent quadratic torsion term. Because $F$ multiplies the quadratic torsion terms contained in the curvature scalar, the effective coefficient is $F+\kappa$ and depends on the field. For $\kappa\to\infty$ the torsion decouples and the theory is the metric scalar--tensor theory with the same $F$, and for $\kappa=0$ it is pure Einstein--Cartan gravity. The difference between the two isolates the effect of torsion and can be computed exactly.

Several ingredients of the construction are known and are not claimed as new. The Einstein-frame form of non-minimally coupled scalar--tensor theories is standard \cite{Futamase1989,Fakir1990,Faraoni1999}, its Einstein--Cartan and Palatini versions are known \cite{Bauer2008,Shaposhnikov2020,Langvik2021}, the conformal covariance of the two-dimensional Laplace--Beltrami operator is a general geometric fact, interference between positive and negative frequency packets is a known source of Bohmian bounces in massless-scalar and stiff-matter minisuperspaces \cite{Colistete2000,PintoNeto2013,Peter2007}, Bessel-function solutions of minisuperspace models with exponential potentials are of a familiar type, and the blue spectrum of a single ekpyrotic field is well established \cite{Lyth2002,Finelli2002,Lehners2008}. What we add, to our knowledge, is the following. The family of models with the field-dependent coefficient $F+\kappa$ gives an exact expression for the part of the kinetic function that is due to torsion, $K-K_{\rm metric}=-3F'^2/[2F(F+\kappa)]$, and shows that the minisuperspace curvature, which might be taken as a signal of torsion, vanishes for non-constant couplings and is present without torsion. The frame map yields closed-form statements about Bohmian bounces in the Jordan frame, namely the exact node structure of a two-packet state, a threshold coupling $\xi_c$ above which the Jordan-frame bounce splits, and torsion-dependent late-time exponents. The steep exponential potential receives an exact analysis of the flux and turning points of single-momentum states. For the ekpyrotic scenario we compute how torsion changes the background, the Jordan-frame contraction and the tilt of the entropy mode in an entropic two-field extension, and we compare these with the corresponding inflationary quantities. We also compute the transfer of tensor perturbations through the Bohmian bounce.

The results are organised as follows. Section~\ref{sec:formalism} defines the model, eliminates the torsion, establishes the structure of the minisuperspace metric including the curvature criterion and its counterexamples, and treats the quantisation and the Bohmian formulation. Section~\ref{sec:massless} treats the massless scalar, the bounce, its Jordan-frame form and the transfer of tensor modes. Section~\ref{sec:exp} treats the exponential potentials, Sec.~\ref{sec:pert} the linear perturbations and inflation, and Sec.~\ref{sec:ekpyrosis} ekpyrotic contraction and its comparison with inflation. Section~\ref{sec:discussion} discusses the implications of the results. Appendix~\ref{app:RC} gives the Riemann--Cartan decomposition and Appendix~\ref{app:num} documents the numerical methods.

\section{Model, minisuperspace geometry and quantisation}
\label{sec:formalism}

We begin by fixing the conventions of the Riemann--Cartan geometry and the action. We use signature $(-,+,+,+)$ and $\hbar$ is kept explicit. The connection $\Gamma^\lambda{}_{\mu\nu}$ is metric
compatible, with covariant derivative $\nabla_\mu V^\lambda=\partial_\mu V^\lambda+\Gamma^\lambda{}_{\mu\nu}V^\nu$,
torsion $T^\lambda{}_{\mu\nu}=\Gamma^\lambda{}_{\mu\nu}-\Gamma^\lambda{}_{\nu\mu}$, and curvature
$R^\rho{}_{\sigma\mu\nu}=\partial_\mu\Gamma^\rho{}_{\nu\sigma}-\partial_\nu\Gamma^\rho{}_{\mu\sigma}+\Gamma^\rho{}_{\mu\lambda}\Gamma^\lambda{}_{\nu\sigma}-\Gamma^\rho{}_{\nu\lambda}\Gamma^\lambda{}_{\mu\sigma}$,
with $R(\Gamma)=g^{\sigma\nu}R^\rho{}_{\sigma\rho\nu}$. The torsion is decomposed into its trace
$T_\mu=T^\nu{}_{\mu\nu}$, its axial vector $S^\mu=\epsilon^{\mu\alpha\beta\gamma}T_{\alpha\beta\gamma}$ and a
tensor part $q_{\lambda\mu\nu}$. With these conventions the curvature scalar decomposes as
\begin{equation}
R(\Gamma)=\tilde R+2\tilde\nabla_\mu T^\mu-\frac23\,T_\mu T^\mu+\frac1{24}\,S_\mu S^\mu+c_q\,q_{\lambda\mu\nu}q^{\lambda\mu\nu},
\label{eq:decomp}
\end{equation}
where a tilde denotes the Levi-Civita quantities and $c_q$ is a numerical constant. The coefficients of the first three
torsion terms are derived in Appendix~\ref{app:RC} for the homogeneous and isotropic sector, which is all that is needed here.
Spatial homogeneity and isotropy admit only a time-dependent trace component and a time-dependent axial component, and the
tensor part vanishes identically \cite{Tsamparlis1979}.

The model is
\begin{equation}
S=\int\dd^4x\,e\left[\frac{\MP^2}{2}F(\phi)R(\Gamma)-\frac12 g^{\mu\nu}\partial_\mu\phi\,\partial_\nu\phi-V(\phi)
-\frac{\MP^2}{3}\kappa\,T_\mu T^\mu+\frac{\MP^2}{48}\kappa_S\,S_\mu S^\mu\right],
\label{eq:action}
\end{equation}
where $e=\sqrt{-g}$, $\MP$ is the reduced Planck mass, $F(\phi)>0$ is dimensionless, and $\kappa$ and $\kappa_S$ are
dimensionless constants. The last two terms are algebraic torsion terms that are allowed by the symmetries and are
not generated by $R(\Gamma)$. The case $\kappa=\kappa_S=0$ is the Einstein--Cartan theory with a non-minimally
coupled scalar. The model \eqref{eq:action} is accurately described as a metric-compatible Riemann--Cartan
scalar--tensor theory with algebraic quadratic torsion terms. Because $F$ multiplies the quadratic torsion terms
contained in $R(\Gamma)$, the effective coefficients of the quadratic torsion forms are $F+\kappa$ and $F+\kappa_S$,
which depend on the field. A constant effective coefficient would not follow from \eqref{eq:action}.

Henceforth we use units $\MP=1$ ($8\pi G=1$), so that $\phi$ and $V$ are measured in units of $\MP$ and $\MP^4$. In these units the
standard example is $F(\phi)=1+\xi\phi^2$ with $\xi>0$ dimensionless.

To reduce the model to minisuperspace we take the spatially flat metric
\begin{equation}
\dd s^2=-N(t)^2\dd t^2+a(t)^2\delta_{ij}\dd x^i\dd x^j,\qquad \phi=\phi(t),
\label{eq:flrw}
\end{equation}
and the most general homogeneous and isotropic torsion, $T_\mu=(u(t),0,0,0)$ and $S_\mu=(v(t),0,0,0)$. Then
$T_\mu T^\mu=-u^2/N^2$, $S_\mu S^\mu=-v^2/N^2$ and $\sqrt{-g}\,\tilde\nabla_\mu T^\mu=-\frac{\dd}{\dd t}(a^3u/N)$.
The comoving volume of the spatial sections is set to unity. Integrating the term $F\tilde\nabla_\mu T^\mu$ by parts, and discarding a boundary term,
the Lagrangian is
\begin{equation}
L=-\frac{3Fa\dot a^2}{N}-\frac{3a^2F'\dot a\dot\phi}{N}+\frac{a^3\dot\phi^2}{2N}-Na^3V
+\frac{a^3}{N}\left[\frac{F+\kappa}{3}\,u^2+F'\dot\phi\,u\right]-\frac{a^3}{48N}(F+\kappa_S)\,v^2 ,
\label{eq:Lfull}
\end{equation}
where the gravitational part is the standard reduction of $\frac12F\tilde R$, a prime denotes $\dd/\dd\phi$, and the dot is $\dd/\dd t$.
Neither $u$ nor $v$ carries a time derivative, so both are auxiliary. The equation for $v$ is homogeneous, and for
$F+\kappa_S\neq0$ it gives $v=0$. The equation for $u$ is
\begin{equation}
\frac{2}{3}(F+\kappa)\,u+F'\dot\phi=0\quad\Longrightarrow\quad u_\star=-\frac{3F'\dot\phi}{2(F+\kappa)} .
\label{eq:ustar}
\end{equation}
This solution requires $\Theta\equiv F+\kappa\neq0$, and we assume $\Theta>0$ throughout, which holds for $\kappa\geq0$. Substituting
\eqref{eq:ustar} and $v=0$ into \eqref{eq:Lfull} gives
\begin{equation}
L_{\rm eff}=\frac{1}{2N}G_{AB}(a,\phi)\,\dot q^A\dot q^B-Na^3V(\phi),\qquad q^A=(a,\phi),
\label{eq:Leff}
\end{equation}
with the minisuperspace metric
\begin{equation}
G_{aa}=-6aF,\qquad G_{a\phi}=-3a^2F',\qquad G_{\phi\phi}=a^3\,\Omega(\phi),\qquad
\Omega(\phi)=1-\frac{3F'^2}{2(F+\kappa)} .
\label{eq:GAB}
\end{equation}
The elimination of $u$ is exact, and no expansion in a small parameter has been used. The same algebraic elimination applied
to the covariant action gives $T_\mu=-\tfrac32 F'\partial_\mu\phi/\Theta$ for arbitrary inhomogeneous configurations, so the
effective four-dimensional theory is the metric theory
\begin{equation}
S_{\rm eff}=\int\dd^4x\sqrt{-g}\left[\frac12F\tilde R-\frac12\,\Omega(\phi)\,g^{\mu\nu}\partial_\mu\phi\,\partial_\nu\phi-V(\phi)\right],
\label{eq:Seff}
\end{equation}
and $\kappa\to\infty$ gives $\Omega\to1$, the torsion-free scalar--tensor theory with the same $F$. This is used in
Sec.~\ref{sec:pert}.

For $F=1$ one has $F'=0$ and torsion is not sourced, so \eqref{eq:Leff} is the minimally coupled minisuperspace
Lagrangian. The determinant is $\det G=-a^4\,[6F\Omega+9F'^2]=-6a^4F^2K$, with
\begin{equation}
K(\phi)\equiv\frac{\Omega}{F}+\frac{3F'^2}{2F^2}=\frac1F+\frac{3\kappa F'^2}{2F^2(F+\kappa)} .
\label{eq:K}
\end{equation}
The metric $G_{AB}$ is therefore Lorentzian if and only if $K>0$. This is the condition that the scalar degree of freedom is
not a ghost in the Einstein frame (Sec.~\ref{sec:pert}), and it holds automatically for $F>0$ and $\kappa\geq0$.
The limiting cases are
\begin{equation}
K_{\rm EC}=\frac1F\quad(\kappa=0),\qquad K_{\rm metric}=\frac1F+\frac{3F'^2}{2F^2}\quad(\kappa\to\infty),
\qquad K-K_{\rm metric}=-\frac{3F'^2}{2F(F+\kappa)} .
\label{eq:Klimits}
\end{equation}
The last expression is the complete effect of torsion on the kinetic function. For $F=1+\xi\phi^2$ it equals
$-6\xi^2\phi^2/[F(F+\kappa)]$, which is of second order in $\xi$. Torsion therefore influences the dynamics only at
second order in the non-minimal coupling at small field, and at order one when $\xi\phi^2\gtrsim1$ (Sec.~\ref{sec:massless}).
The ordinary non-minimal coupling already contributes at first order in $\xi$ and must not be confused with it.

Having eliminated the torsion, we examine the structure of the reduced minisuperspace metric, which determines most of what follows. We define the Einstein-frame variables
\begin{equation}
\tilde a=\sqrt{F(\phi)}\,a,\qquad \tilde N=\sqrt{F(\phi)}\,N,\qquad \chi(\phi)=\int^\phi\!\sqrt{K(\phi')}\,\dd\phi' ,
\label{eq:einstein-map}
\end{equation}
on any interval where $F>0$, $\Theta=F+\kappa>0$ and $K>0$. Because $\tilde a=\sqrt F\,a$ depends on $\phi$, one has $\dd\tilde a=\sqrt F\,\dd a+\tfrac12aF'F^{-1/2}\dd\phi$. Substituting this into $-6\tilde a\,\dd\tilde a^2+\tilde a^3K\dd\phi^2$ and using the definition \eqref{eq:K} of $K$ reproduces $\sqrt F\,G_{AB}\dd q^A\dd q^B$ term by term: the $\dd a^2$ coefficient is $-6aF^{3/2}$, the $\dd a\,\dd\phi$ coefficient is $-3\sqrt F\,a^2F'$, and the $\dd\phi^2$ coefficient is $\sqrt F\,a^3\Omega$. The potential term follows from $Na^3V=(\tilde N/\sqrt F)\,\tilde a^3F^{-3/2}V$. We also verified the identity of the quadratic forms symbolically for general $F(\phi)$ and $\kappa$. The Lagrangian \eqref{eq:Leff} is therefore, exactly and without any expansion,
\begin{equation}
L_{\rm eff}=-\frac{3\tilde a\,\dot{\tilde a}^2}{\tilde N}+\frac{\tilde a^3\dot\chi^2}{2\tilde N}-\tilde N\tilde a^3\,U(\chi),\qquad U=\frac{V}{F^2},
\label{eq:Leinstein}
\end{equation}
which means that $\sqrt F\,G_{AB}\,\dd q^A\dd q^B=\tilde G_{AB}\,\dd\tilde q^A\dd\tilde q^B$ with the flat metric $\tilde G=-6\tilde a\,\dd\tilde a^2+\tilde a^3\dd\chi^2=\tilde a^3(-6\,\dd\tilde\alpha^2+\dd\chi^2)$ and $\tilde\alpha=\ln\tilde a$. For every $\kappa$ the homogeneous sector of the torsionful model is thus classically equivalent to general relativity with one minimally coupled scalar $\chi$ and the potential $U=V/F^2$, and the only trace of $\kappa$ is the function $\chi(\phi)$ through $K(\phi)$ in \eqref{eq:K}. For the quadratic coupling with $\kappa=0$ one finds in closed form
\begin{equation}
\chi=\frac{1}{\sqrt\xi}\,\mathrm{arcsinh}\!\big(\sqrt\xi\,\phi\big),\qquad \phi=\frac{\sinh(\sqrt\xi\chi)}{\sqrt\xi},\qquad F=\cosh^2(\sqrt\xi\,\chi),
\label{eq:EC-map}
\end{equation}
whereas in the torsion-free case $\chi(\phi)$ is given by the integral of $[1/F+6\xi^2\phi^2/F^2]^{1/2}$.

The curvature of the minisuperspace metric follows from the same structure. Since $G_{AB}=F^{-1/2}\,\tilde G_{AB}$ with $\tilde G$ flat, and since $\sqrt F$ depends on $\phi$, hence on $\chi$, only, we can use the two-dimensional relation $R[\ee^{2\omega}\tilde G]=\ee^{-2\omega}(\tilde R-2\tilde\Box\omega)$. With $\omega=-\tfrac14\ln F$, $\tilde R=0$ and $\tilde\Box\omega=\tilde a^{-3}\omega_{,\chi\chi}$ for $\tilde G=\tilde a^3(-6\dd\tilde\alpha^2+\dd\chi^2)$, one obtains $R[G]=\tfrac12F^{1/2}\tilde a^{-3}(\ln F)_{,\chi\chi}$, which becomes
\begin{equation}
R[G]=\frac{(\ln F)_{,\chi\chi}}{2a^3F},\qquad (\cdot)_{,\chi}\equiv K^{-1/2}\frac{\dd}{\dd\phi},
\label{eq:RG}
\end{equation}
because $\tilde a^{3}=F^{3/2}a^3$. A direct computation of the Ricci scalar of \eqref{eq:GAB} gives the equivalent expression in terms of the coupling and its derivatives,
\begin{equation}
R[G]=\frac{2\Theta^2\,(2FF''-F'^2)+3\kappa F'^4}{2a^3\,(2F\Theta+3\kappa F'^2)^2},
\label{eq:RGexplicit}
\end{equation}
and we checked symbolically that the two expressions agree. It follows that $R[G]$ vanishes identically on an interval if and only if $\ln F$ is affine in $\chi$, that is $F=F_0\,\ee^{\lambda\chi}$. Inserting $F'=\lambda\sqrt K\,F$ into \eqref{eq:K} gives
\begin{equation}
F'^2=\frac{\lambda^2F\,(F+\kappa)}{F+\kappa-\tfrac32\lambda^2\kappa},
\label{eq:flatODE}
\end{equation}
a first-order separable equation whose denominator is positive for $|\lambda|$ small enough, so that non-constant solutions exist for every $\kappa\ge0$. For $\kappa=0$ the equation reduces to $F'^2=\lambda^2F$ with solutions $F=(\lambda\phi/2+c)^2$, and for $\kappa\to\infty$ to $F'^2=\lambda^2F/(1-\tfrac32\lambda^2)$, whose solutions are again quadratic polynomials, for example $F=\xi\phi^2$. Non-vanishing curvature is therefore neither necessary nor sufficient for the presence of torsion: it is present for $\kappa\to\infty$ and absent for non-constant couplings of the form \eqref{eq:flatODE} at any $\kappa$.

For $F=1+\xi\phi^2$ and small $\xi$ one finds $R[G]=\xi/a^3+O(\xi^2)$, and this leading term is present for $\kappa\to\infty$, so the first-order curvature of the minisuperspace metric is a property of the non-minimal coupling alone. Torsion contributes to $R[G]$ only through the dependence of $K$ on $\kappa$, which enters at second order in $\xi$. The scalar $R[G]$ is also not invariant under changes of lapse, because $N\to fN$ changes $G\to G/f$ and $R[G/f]\neq R[G]/f$ in general, and \eqref{eq:RG} shows that its non-zero value reflects the conformal factor $F^{-1/2}$ that separates the Jordan lapse from the Einstein lapse. A lapse-independent geometric object is the Jacobi metric $U\,G_{AB}$, which is degenerate for $V=0$.

With the geometry of the reduced configuration space established, we turn to the quantisation and the Bohmian formulation. To quantise the reduced model we start from the constraint. Varying the lapse in \eqref{eq:Leff} gives the Hamiltonian constraint $\mathcal H=\tfrac12G^{AB}p_Ap_B+U_J=0$ with $U_J=a^3V$.
No torsion variable remains, since $u$ was eliminated before the Legendre transformation. We quantise with the
Laplace--Beltrami ordering \cite{HawkingPage1986,Halliwell1988},
\begin{equation}
\left[-\frac{\hbar^2}{2}\Box_G+a^3V(\phi)\right]\Psi(a,\phi)=0,\qquad
\Box_G\Psi=\frac{1}{\sqrt{|G|}}\,\partial_A\!\left(\sqrt{|G|}\,G^{AB}\partial_B\Psi\right),
\label{eq:wdw}
\end{equation}
where $\Box_G$ is built from the Levi-Civita connection of $G_{AB}$. Torsion enters only through $G_{AB}$ and no first-order term
is added. Laplace--Beltrami ordering is one of several possible choices, and an alternative is the addition of a term
proportional to $\hbar^2R[G]\Psi$; we return to this below.

The Laplace--Beltrami Wheeler--DeWitt equation has a property that follows directly from the structure established above. In two dimensions the Laplace--Beltrami operator obeys $\Box_{\lambda^2 g}=\lambda^{-2}\Box_g$, so with $\tilde G=F^{1/2}G$ one has $\Box_{\tilde G}=F^{-1/2}\Box_G$. In addition $\tilde a^3U=F^{3/2}a^3V/F^2=F^{-1/2}a^3V$, and therefore the operator in
\begin{equation}
\left[-\frac{\hbar^2}{2}\Box_{\tilde G}+\tilde a^3\,U(\chi)\right]\Psi=0
\label{eq:wdwE}
\end{equation}
equals $F^{-1/2}$ times the operator in \eqref{eq:wdw}. A function $\Psi$ solves \eqref{eq:wdw} if and only if it solves \eqref{eq:wdwE}. The same argument applies to the lapse in general: for any positive function $f(q)$ the replacement $N\to fN$ maps $G\to G/f$ and $U_J\to fU_J$, and the Wheeler--DeWitt operator becomes $f$ times itself, so the space of solutions does not depend on the choice of lapse. The Bohmian velocity $\dot q^A=G^{AB}\partial_BS$ is rescaled by $f$ under $G\to G/f$, so the integral curves in configuration space are unchanged and only their parametrisation depends on the lapse. We verified the operator identity $\Box_G\Psi=F^{1/2}\Box_{\tilde G}\Psi$ symbolically and numerically for test functions.

This removes the lapse ambiguity that is present in higher-dimensional minisuperspace models; it is special to two
dimensions, where the conformal coupling constant vanishes. It also shows that the Wheeler--DeWitt equation of the torsionful
model with Laplace--Beltrami ordering is identical to that of a minimally coupled scalar $\chi$ with potential
$U(\chi)$. The ordering ambiguity is therefore reduced to the possible curvature term $\zeta\hbar^2R[G]\Psi$. Because
$R[G]$ involves derivatives of the conformal factor and does not rescale homogeneously, a non-zero $\zeta$ breaks the frame covariance, and the Einstein-frame
equation would then differ from the Jordan-frame equation. We set $\zeta=0$, which is the choice that makes the quantum theory independent of the
frame.

The Bohmian formulation follows from writing $\Psi=\mathcal R\,\ee^{\ii S/\hbar}$ with real $\mathcal R$ and $S$ gives, exactly and without an expansion in $\hbar$,
\begin{align}
\tfrac12G^{AB}\partial_AS\,\partial_BS+a^3V+\mathcal Q&=0,\qquad \mathcal Q=-\frac{\hbar^2}{2\mathcal R}\Box_G\mathcal R,\label{eq:qhj}\\
\nabla_AJ^A&=0,\qquad J^A=\mathcal R^2G^{AB}\partial_BS,\label{eq:cont}
\end{align}
and the guidance equations are $\dot q^A=G^{AB}\partial_BS$. The semiclassical limit is the regime $|\mathcal Q|\ll|a^3V|$ and $|\mathcal Q|\ll|G^{AB}S_{,A}S_{,B}|$ in which
\eqref{eq:qhj} reduces to the classical Hamilton--Jacobi equation. The usual expansion $S=S_0+\hbar S_1+\dots$ produces at order
$\hbar$ the cross term $G^{AB}S_{0,A}S_{1,B}$, which must be combined with \eqref{eq:cont}; the equivalent and cleaner statement is that
$\Psi=\ee^{\ii S_0/\hbar}(C_0+\hbar C_1+\dots)$ with $\nabla_A(C_0^2G^{AB}S_{0,B})=0$ at the first non-trivial order. The exact
decomposition above does not depend on the expansion and is the one used below.

Because \eqref{eq:wdw} is hyperbolic, $\mathcal R^2$ is not a positive conserved density on configuration space. What is conserved is the
current \eqref{eq:cont}. For a slice $\Sigma$ of constant $\tilde\alpha$, which is spacelike for $\tilde G$, the flux density is
$\sqrt{|\tilde G|}\,J^{\tilde\alpha}=-\tfrac{1}{\sqrt6}\mathcal R^2S_{,\tilde\alpha}$ and is positive on a branch with definite sign of $S_{,\tilde\alpha}$.
We define Bohmian ensembles by weighting the initial positions on $\Sigma$ with $\mathcal R^2|S_{,\tilde\alpha}|$. Probabilities defined in this way are
conserved along the flow. No claim is made about a physical inner product or about boundary conditions at $a=0$; the states
considered below are specified by their form on the incoming branch, and flux conservation is used instead of a norm.

\section{The massless scalar: exact solutions and Bohmian bounces}
\label{sec:massless}

We now consider the massless scalar. For $V=0$ one has $U=0$ in every frame, and by Eq.~\eqref{eq:wdwE} the Wheeler--DeWitt equation is exactly
\begin{equation}
\Psi_{,\tilde\alpha\tilde\alpha}=6\,\Psi_{,\chi\chi},
\label{eq:wave}
\end{equation}
for all $F$, all $\kappa$, and all values of $\xi$ without any expansion. The general solution is
$\Psi=f(x_+)+g(x_-)$ with $x_\pm=\chi\pm\sqrt6\,\tilde\alpha$. We take the two-packet state ($\hbar=1$)
\begin{equation}
\Psi=\mathsf F(x_+)+\mathsf F(x_-),\qquad \mathsf F(x)=\sqrt{\pi/d}\,\exp\!\Big[\ii\bar kx-\frac{x^2}{4d}\Big],
\label{eq:twopacket}
\end{equation}
with real $\bar k>0$ and $d>0$. A single packet has $S=\bar kx_\pm$, a constant guidance momentum, and never turns. Since
$\mathsf F(-x)=\mathsf F(x)^*$, one has $\Psi(\tilde\alpha,-\chi)=\Psi(\tilde\alpha,\chi)^*$, so $S$ is odd in $\chi$. States of this type are Gaussian superpositions of positive and negative frequency modes of the massless minimally coupled scalar, which were used to construct Bohmian bounces in the Einstein-frame problem in Ref.~\cite{Colistete2000}. In the present setting they supply a definite state on which the frame map and the dependence on torsion can be analysed exactly. Units are $\hbar=8\pi G=V_0=1$, where $V_0$ is
the comoving volume.

The guidance equations in Einstein-frame time, $\dd\tau=\tilde a^{-3}\dd\tilde t$ with $\tilde t$ the Einstein cosmic time, are
\begin{equation}
\tilde\alpha'=-\tfrac16S_{,\tilde\alpha},\qquad \chi'=S_{,\chi}.
\label{eq:guid}
\end{equation}

The structure of the state on and near the symmetry line $\chi=0$ can be determined exactly. A zero of $\Psi$ requires $|\mathsf F(x_+)|=|\mathsf F(x_-)|$, hence $x_+^2=x_-^2$, hence $\chi\tilde\alpha=0$. At $\tilde\alpha=0$ one has $x_+=x_-$ and $\Psi=2\mathsf F\neq0$. At $\chi=0$ one has $\Psi=2\,\mathrm{Re}\,\mathsf F(\sqrt6\tilde\alpha)=2\sqrt{\pi/d}\,\ee^{-3\tilde\alpha^2/2d}\cos(\sqrt6\bar k\tilde\alpha)$. The zeros of $\Psi$ are therefore exactly the isolated points $(\tilde\alpha_n,0)$ with $\sqrt6\,\bar k\,\tilde\alpha_n=\pi/2+n\pi$, $n\in\mathbb Z$. Near $\chi=0$ we write $\Psi=A(\tilde\alpha)+\ii\chi B(\tilde\alpha)+O(\chi^2)$ with $A$ and $B$ real, which uses $\mathsf F'(-x)=-\overline{\mathsf F'(x)}$ for the derivative with respect to $\chi$. Then $S_{,\tilde\alpha}=\mathrm{Im}(\Psi_{,\tilde\alpha}/\Psi)=0$ on the line, $S_{,\chi}=B/A$, and $S_{,\tilde\alpha\chi}=(B/A)'$. Evaluating $B/A=\mathrm{Im}\,\mathsf F'/\mathrm{Re}\,\mathsf F$ gives the guidance momentum along the symmetry line,
\begin{equation}
p(\tilde\alpha)\equiv S_{,\chi}(\tilde\alpha,0)=\bar k-\frac{\sqrt6\,\tilde\alpha}{2d}\tan\!\big(\sqrt6\,\bar k\tilde\alpha\big),
\qquad S_{,\tilde\alpha\chi}(\tilde\alpha,0)=p'(\tilde\alpha).
\label{eq:p}
\end{equation}

\begin{figure}[t]
\centering
\includegraphics[width=\textwidth]{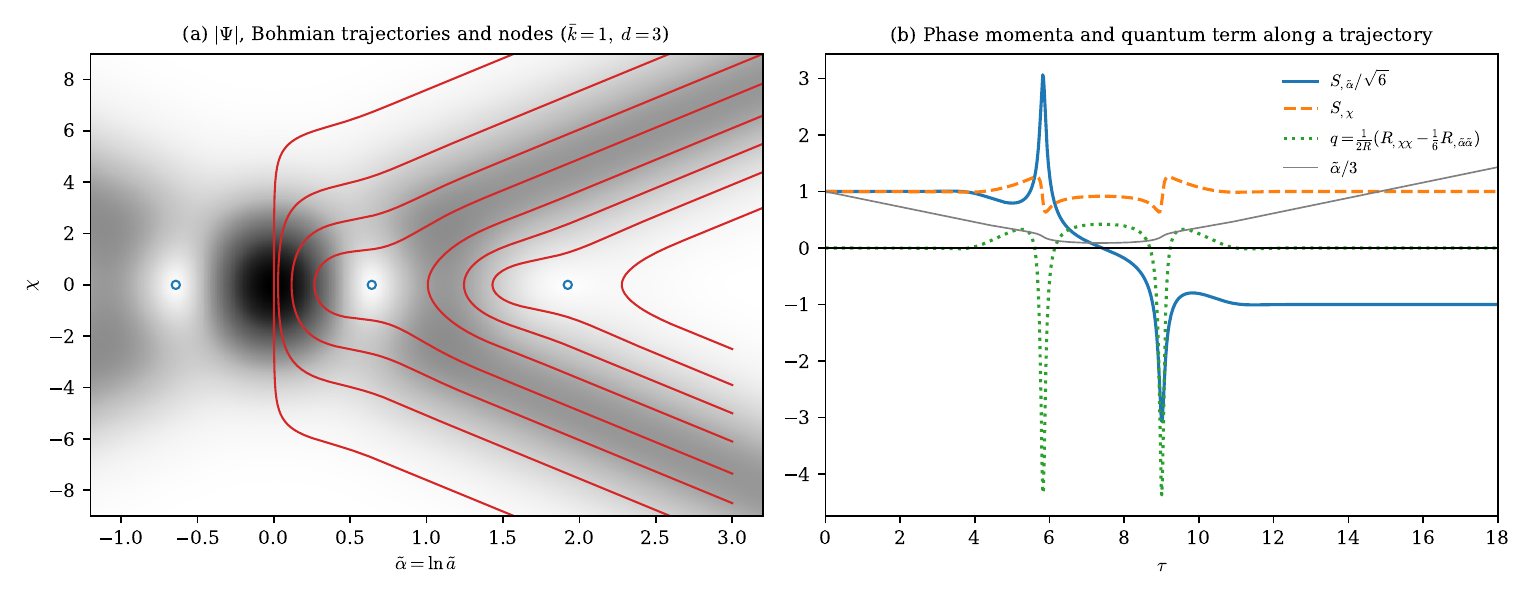}
\caption{Massless scalar in the Einstein frame for $\bar k=1$, $d=3$, $\hbar=8\pi G=V_0=1$.
(a) Grey scale: $|\Psi|$ in the $(\tilde\alpha,\chi)$ plane. Red: Bohmian trajectories launched at $\tilde\alpha=3$. Circles: the nodes of Eq.~\eqref{eq:p}.
(b) $S_{,\tilde\alpha}/\sqrt6$, $S_{,\chi}$, the quantum term $\mathsf q$ and $\tilde\alpha/3$ along the trajectory launched at $x_+=0$.
The spikes occur where the trajectory passes near the nodes at $\tilde\alpha=0.64$ and $1.92$.}
\label{fig:einstein}
\end{figure}

The nodes at $\tilde\alpha_n>0$ are of the same size as the bounce scale for the parameters used here, and trajectories pass at distances of the order of the packet
width from them (Fig.~\ref{fig:einstein}a). Bohmian non-crossing makes each node a vortex of the flow, and trajectories on the two sides of a node
turn at different values of $\tilde\alpha$.

The occurrence of bounces is governed by the behaviour on the symmetry line. Every trajectory that crosses $\chi=0$ at a non-nodal point $\tilde\alpha_b$ has $\tilde\alpha'=0$ there, because $S_{,\tilde\alpha}=0$ on this line. Differentiating $\tilde\alpha'=-S_{,\tilde\alpha}/6$ along the flow gives $\tilde\alpha''=-\tfrac16(S_{,\tilde\alpha\tilde\alpha}\tilde\alpha'+S_{,\tilde\alpha\chi}\chi')$, and at the crossing $\tilde\alpha'=0$ and $\chi'=p$, so that
\begin{equation}
\tilde\alpha''=-\tfrac16\,p(\tilde\alpha_b)\,p'(\tilde\alpha_b).
\label{eq:alpha2}
\end{equation}
The crossing is therefore a minimum of the Einstein-frame scale factor, a bounce, if and only if $p\,p'<0$ at $\tilde\alpha_b$.

Fig.~\ref{fig:einstein} shows trajectories launched on the incoming branch. We launched 41 trajectories on the slice $\tilde\alpha=3$,
uniformly spaced over $\pm2.5\sqrt{2d}$ around $x_+=0$, for $\bar k=1$, $d=3$, with weights $\mathcal R^2|S_{,\tilde\alpha}|$ (Sec.~\ref{sec:formalism}).
All of them turn around. The flux-weighted bounce locations have median $\tilde\alpha_b=0.27$ ($a_b=1.30$) and a 16 to 84 per cent range
$0.06\le\tilde\alpha_b\le1.14$ ($1.07\le a_b\le3.14$). The statement is numerical and applies to this sample; the hyperbolic form of
\eqref{eq:wave} does not provide a probability measure in which a fraction of ``all trajectories'' can be defined without
specifying the slice. On both asymptotic branches $S_{,\tilde\alpha}\to\pm\sqrt6\,\bar k$ and $S_{,\chi}\to\bar k$, so the
trajectories approach the classical stiff-fluid solutions on both sides. Along every trajectory the Hamilton--Jacobi
equation \eqref{eq:qhj} was checked numerically; for $V=0$ it reads $\tfrac12S_{,\chi}^2-\tfrac1{12}S_{,\tilde\alpha}^2=\mathsf q$ with
$\mathsf q=(2\mathcal R)^{-1}(\mathcal R_{,\chi\chi}-\tfrac16\mathcal R_{,\tilde\alpha\tilde\alpha})$, and Fig.~\ref{fig:einstein}b shows that the
bounce coincides with the region where $\mathsf q$ is non-zero.

The scale factor of the bounce has no meaning in isolation, because the comoving volume sets the unit of $a$. An invariant measure is the energy density of the scalar at the bounce,
$\rho_b=p(\tilde\alpha_b)^2/(2\tilde a_b^6)$ in the units above, and the ratio $\tilde a_b/\tilde a_*$ with $\tilde a_*$ the scale at which the centres of the two packets intersect ($\tilde a_*=1$ here).
Table~\ref{tab:bounce} lists these quantities for the trajectory launched at $x_+=0$.

\begin{table}[t]
\centering
\caption{Bounce of the trajectory launched at $x_+=0$, $\tilde\alpha=3$, for several packet parameters (units $\hbar=8\pi G=V_0=1$): bounce location,
the density $\rho_b$, and the threshold coupling $\xi_c=-p'/(6p)$ of Eq.~\eqref{eq:xic}.}
\label{tab:bounce}
\begin{tabular}{ccccccc}
\toprule
$\bar k$ & $d$ & $\tilde\alpha_b$ & $\tilde a_b/\tilde a_*$ & $p(\tilde\alpha_b)$ & $\rho_b$ & $\xi_c$\\
\midrule
1   & 1 & 0.138 & 1.148 & 0.941 & 0.193 & 0.159\\
1   & 3 & 0.266 & 1.305 & 0.917 & 0.0853 & 0.133\\
1   & 8 & 0.970 & 2.638 & 1.143 & 0.00194 & 0.081\\
0.5 & 3 & 0.236 & 1.267 & 0.471 & 0.0269 & 0.088\\
2   & 3 & 0.552 & 1.737 & 2.105 & 0.0807 & 0.091\\
\bottomrule
\end{tabular}
\end{table}

The Einstein-frame results translate into the Jordan frame as follows. The Jordan-frame scale factor along a trajectory is $a=\tilde a/\sqrt{F(\phi(\chi))}$, and the Jordan cosmic time satisfies
$\dd t_J=\dd\tilde t/\sqrt F=a^3F\,\dd\tau$. Because the flow lines are frame independent (Eq.~\eqref{eq:wdwE}), the trajectory
in the Jordan frame is the same curve in the $(\tilde\alpha,\chi)$ plane mapped through $\phi(\chi)$. Whether the Jordan
scale factor bounces is a separate question, because $a$ and $\tilde a$ differ by a field-dependent factor. For $F=1+\xi\phi^2$ one has $F(0)=1$, $F'(0)=0$,
$K(0)=1$, and therefore for every $\kappa$
\begin{equation}
(\ln F)_{,\chi\chi}\big|_{\chi=0}=\frac{F''(0)}{F(0)K(0)}=2\xi .
\label{eq:lnFcc}
\end{equation}

The Jordan-frame scale factor $\ln a=\tilde\alpha-\tfrac12\ln F$ has a critical point at the Einstein-frame bounce event for every $\xi$ and $\kappa$. At the event $\tilde\alpha'=0$, and $\dd\ln F/\dd\tau=(\ln F)_{,\chi}\chi'$ vanishes at $\chi=0$ because $F$ is even in $\chi$. The second derivative along the flow is $\tilde\alpha''-\tfrac12(\ln F)_{,\chi\chi}\chi'^2$ with $\chi'=p$, and inserting \eqref{eq:alpha2} and \eqref{eq:lnFcc} gives
\begin{equation}
\frac{\dd^2\ln a}{\dd\tau^2}\bigg|_b=-\frac16\,p\,p'-\xi\,p^2 .
\label{eq:jordan2}
\end{equation}
This expression does not depend on $\kappa$, and the sign of the second derivative at a critical point does not depend on the parametrisation of time. The Jordan-frame scale factor therefore has a minimum at the event if and only if
\begin{equation}
\xi<\xi_c(\tilde\alpha_b)\equiv-\frac{p'(\tilde\alpha_b)}{6\,p(\tilde\alpha_b)} .
\label{eq:xic}
\end{equation}
For $\xi>\xi_c$ the event is a local maximum of $a$ flanked by two minima, so that the single bounce is replaced by two bounces separated by a short episode of expansion.

\begin{figure}[t]
\centering
\includegraphics[width=\textwidth]{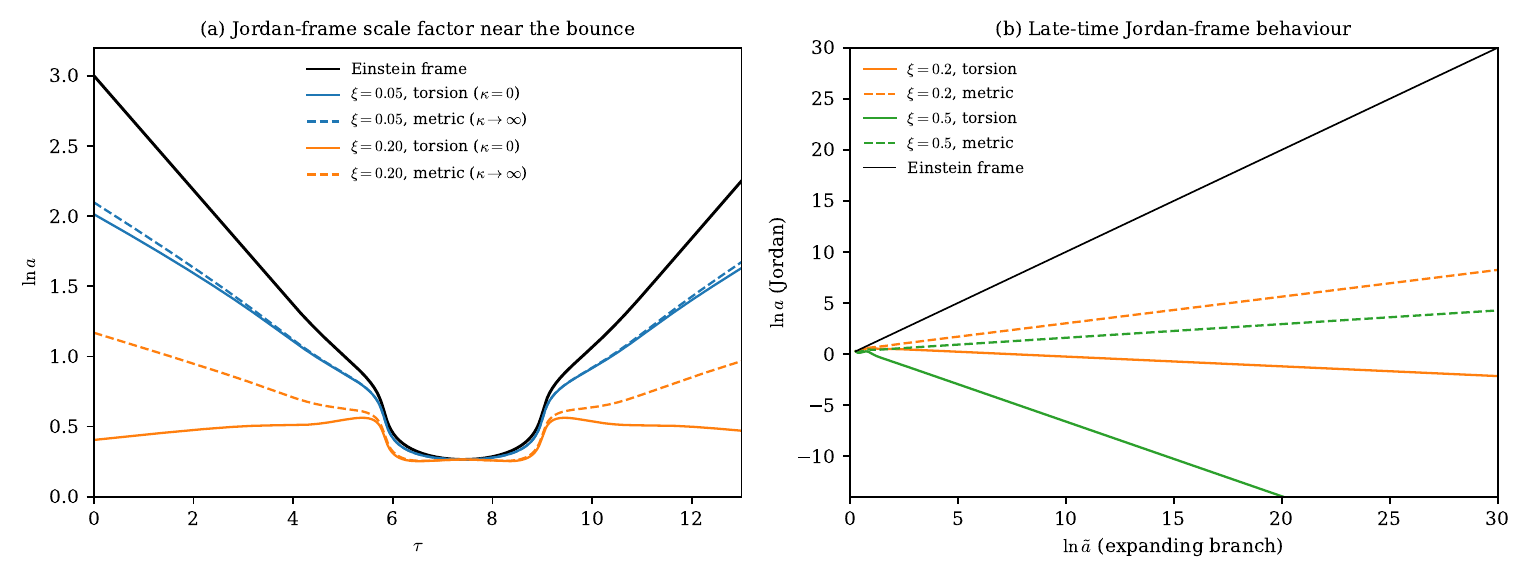}
\caption{Jordan-frame scale factor along the trajectory of Fig.~\ref{fig:einstein}. Solid: $\kappa=0$ (Einstein--Cartan). Dashed: $\kappa\to\infty$ (torsion free).
Black: Einstein frame. (a) Near the bounce for $\xi=0.05$ (below $\xi_c=0.133$) and $\xi=0.2$ (above $\xi_c$, where the bounce has split). The curves for the two
values of $\kappa$ coincide at the bounce event. (b) Expanding branch for $\xi=0.2$ and $0.5$, showing the exponents of Eq.~\eqref{eq:slopes}.}
\label{fig:jordan}
\end{figure}

For the packet $\bar k=1$, $d=3$, the event has $\xi_c=0.133$. In a direct integration of the trajectory with the exact map $\phi(\chi)$, the bounce is a single minimum for $\xi=0.13$ and
is replaced by a local maximum with two flanking minima for $\xi=0.14$, for both $\kappa=0$ and $\kappa\to\infty$ (Fig.~\ref{fig:jordan}a shows $\xi=0.05$ and $0.2$).
The independence of $\kappa$ reflects the fact that torsion modifies $K$ only through terms of order $\phi^2$ times $F'^2$, which vanish to the required order at $\phi=0$. Away from
$\chi=0$ the torsion dependence appears at order $\xi^3\chi^4$: expanding at small $\chi$ gives $\ln a_{\rm EC}-\ln a_{\rm metric}=-\xi^3\chi^4+O(\chi^6)$.

The late-time behaviour differs between the theories with and without torsion. On both asymptotic branches $x_\pm$ is constant, so that $|\chi|=\sqrt6\,\tilde\alpha+\text{const}$. For large $\phi$ and finite $\kappa$ one has $F\simeq\xi\phi^2$ and the second term in \eqref{eq:K} is of order $\phi^{-4}$, so $K\simeq1/(\xi\phi^2)$, $\phi\propto\ee^{\sqrt\xi|\chi|}$ and $\ln F\simeq2\sqrt\xi|\chi|$. For $\kappa\to\infty$ the second term in $K_{\rm metric}$ is $6/\phi^2$, which gives $K\simeq(1+6\xi)/(\xi\phi^2)$ and $\ln F\simeq2\sqrt{\xi/(1+6\xi)}\,|\chi|$. Since $\ln a=\tilde\alpha-\tfrac12\ln F$, on both asymptotic branches $\ln a=s\,\tilde\alpha+\text{const}$ with
\begin{equation}
s=1-\sqrt{6\xi}\quad(\text{any finite }\kappa\ge0),\qquad s_{\rm metric}=1-\sqrt{\frac{6\xi}{1+6\xi}}\quad(\kappa\to\infty).
\label{eq:slopes}
\end{equation}

The numerical trajectory confirms the exponents: for $\xi=0.2$ the slope of $\ln a$ against $\tilde\alpha$ is $-0.0954$ for $\kappa=0,1,10,10^4$ and $0.2615$ for the torsion-free theory.
For $\xi>1/6$ the exponent $1-\sqrt{6\xi}$ is negative for every finite $\kappa$: the Jordan scale factor decreases while the Einstein-frame scale factor
expands. This regime is $\xi\phi^2\gtrsim1$, outside the validity of any expansion at first order in $\xi$, and it is the regime in which torsion is not a small
correction. The two frames are mathematically equivalent, so which of the two scale factors is regarded as physical depends on the
couplings of other matter, which are absent here.

The exact structure above also clarifies the content of an expansion in $\xi$. In a first-order treatment in $\xi$ the wave function, the guidance field and the bounce are all functions of $\xi$ through the metric $G_{AB}$.
The exact structure above shows what such an expansion contains. The wave function and the flow lines do not depend on $\xi$ or $\kappa$ at all in the Einstein variables.
All dependence enters through the map from $(\tilde\alpha,\chi)$ to $(a,\phi)$, and a statement about a shift of the bounce at fixed $\xi$ is a statement about how the state
is specified. If the state is held fixed in the Einstein variables, as here, the bounce event is unchanged and only its depth in $a$ changes, by
\eqref{eq:jordan2}. If instead data are imposed on a surface $\alpha=\text{const}$ in the Jordan variables, the surface corresponds to $\tilde\alpha=\alpha+\tfrac12\ln F(\phi)$, which is not a surface of constant $\tilde\alpha$,
and a shift of the bounce arises from the change of the surface. In either case the shift of the bounce is not a torsion effect at first order in $\xi$.

Beyond the background dynamics, the transfer of long-wavelength perturbations through the bounce can be computed unambiguously for tensor modes. In the Einstein frame, with the Bohmian background $\tilde a(\tau)$ of the trajectory and the conformal time $\dd\eta=\tilde a^2\,\dd\tau$ (from $\dd\eta=\dd\tilde t/\tilde a$ and $\dd\tilde t=\tilde a^3\dd\tau$), the tensor equation $h_{,\eta\eta}+2\tilde a_{,\eta}\tilde a^{-1}h_{,\eta}+k^2h=0$ becomes, exactly,
\begin{equation}
h''+k^2\,\ee^{4\tilde\alpha(\tau)}\,h=0 ,
\label{eq:tensorbounce}
\end{equation}
because the terms in $\tilde\alpha'$ cancel. The tensor amplitude $h$ is frame independent (Sec.~\ref{sec:pert}), so the result applies equally to the Jordan frame. For $k^2\tilde a^4\ll1$ the solution is $h=A+B(\tau-\tau_b)$. we integrated \eqref{eq:tensorbounce} along the trajectory launched at $\tilde\alpha=6$, $x_+=0$ for $\bar k=1$, $d=3$ (bounce at $\tau_b=14.76$, $\tilde\alpha_b=0.2659$), with Bunch--Davies initial data $h=(2k\tilde a^2)^{-1/2}$, $h'=(-\ii k\tilde a^2-\tilde\alpha')h$ at $\tau=0$, where $\tilde a=\ee^{6}$, for $2\times10^{-4}\le k\le3.2\times10^{-3}$. The constants $A$ and $B$ were read off at $\tilde\alpha=1.8$ on the contracting and on the expanding side, where $k^2\tilde a^4\le0.014$.
\begin{table}[t]
\centering
\caption{Long-wavelength tensor modes before and after the Bohmian bounce ($\bar k=1$, $d=3$). $A$ and $B$ are the constant and linear coefficients of $h=A+B(\tau-\tau_b)$.}
\label{tab:tensor}
\begin{tabular}{ccccc}
\toprule
$k$ & $|A|_{\rm in}$ & $|A|_{\rm out}$ & $|B|_{\rm in}$ & $|B|_{\rm out}$\\
\midrule
$2\times10^{-4}$ & 5.389 & 5.389 & 0.5099 & 0.5096\\
$8\times10^{-4}$ & 4.545 & 4.553 & 0.5111 & 0.5064\\
$3.2\times10^{-3}$ & 3.756 & 3.883 & 0.5233 & 0.4628\\
\bottomrule
\end{tabular}
\end{table}
The bounce is nearly transparent: $A$ and $B$ change by less than $0.1\%$ at $k=2\times10^{-4}$ and by at most $3.4\%$ in $A$ and $12\%$ in $B$ at $k=3.2\times10^{-3}$, with the change growing with $k$ as expected for corrections of order $k^2\tilde a^4$. The spectra $k^3|A|^2$ and $k^3|B|^2$ have logarithmic slopes $2.74$ and $3.02$ before the bounce and $2.76$ and $2.94$ after it. These are the blue slopes of the incoming stiff-fluid contraction ($\epsilon=3$, $\nu=0$, for which the pure power law $k^3$ is modified by logarithms), and the bounce does not change them. The tensor spectrum after the bounce is therefore blue, $n_t\simeq2.7$ to $3.0$, and the constant mode $A$ dominates.
For scalar perturbations the standard Mukhanov--Sasaki variable has $\tilde z=\tilde a\sqrt{2\epsilon}$ with $\epsilon=\tilde\chi'^2/(2\tilde H^2)$ diverging at the bounce, where $\tilde H=0$, so a bounce calculation requires the perturbation theory on the Bohmian background and not the classical Mukhanov--Sasaki equation. This calculation is not performed here.

\section{Exponential potentials: classical scaling solutions and exact quantum solutions}
\label{sec:exp}

We now turn to exponential potentials. Take in the Einstein frame $U(\chi)=-V_0\,\ee^{-\lambda\chi}$ with $V_0>0$ and $\lambda>0$. In the Jordan frame this is the potential
$V(\phi)=-V_0F^2\,\ee^{-\lambda\chi(\phi)}$, and for the quadratic coupling with $\kappa=0$,
$V=-V_0(1+\xi\phi^2)^2\,(\sqrt\xi\phi+\sqrt{1+\xi\phi^2})^{-\lambda/\sqrt\xi}$, which tends to $-V_0\ee^{-\lambda\phi}$ for $\xi\to0$.
By Eq.~\eqref{eq:Leinstein} the classical dynamics is that of a minimally coupled scalar with this potential, and the torsion enters only through the map $\chi(\phi)$.

With $\epsilon\equiv-\dot H/H^2$ in the Einstein frame and cosmic time $t$, the exponential potential has the scaling solution
\begin{equation}
a\propto T^{1/\epsilon},\quad T=-t>0,\qquad \epsilon=\frac{\lambda^2}{2},\qquad \chi=\frac{2}{\lambda}\ln T+\chi_0,
\qquad V_0\,\ee^{-\lambda\chi_0}=\frac{\epsilon-3}{\epsilon^2},
\label{eq:scaling}
\end{equation}
where we write $a$ for the Einstein-frame scale factor in this section. To derive it, set $H=-1/(\epsilon T)$. The relation $\dot H=-\dot\chi^2/2$ gives $|\dot\chi|=\sqrt{2/\epsilon}\,/T=2/(\lambda T)$, so
$\chi=(2/\lambda)\ln T+\chi_0$ for a field rolling toward $\chi\to-\infty$ as $T\to0$. The Friedmann equation $3H^2=\dot\chi^2/2+U$ then gives $U=-(\epsilon-3)/(\epsilon^2T^2)$, which equals
$-V_0\ee^{-\lambda\chi_0}/T^2$ and fixes $\chi_0$. The solution exists with real $\chi_0$ if and only if $\epsilon>3$, that is $\lambda>\sqrt6$. The equation of state of the scaling solution is
$w=2\epsilon/3-1=\lambda^2/3-1$, so $\epsilon>3$ is equivalent to $w>1$. This is the ekpyrotic condition, and $\lambda=\sqrt6$ is the marginal stiff-fluid case $w=1$. For $\epsilon>3$ the scaling solution is an attractor of the contracting
phase \cite{Gratton2004,Lehners2008}. A numerical integration of the Einstein-frame equations for $\lambda=3$ ($\epsilon=4.5$) from two different initial conditions gives
$\epsilon_{\rm eff}=-\dot H/H^2\to4.51$ after a few e-folds of contraction, which confirms the attractor.

The quantum problem for the Einstein-frame exponential is as follows. With $\mu=2V_0/\hbar^2$ the Wheeler--DeWitt equation \eqref{eq:wdwE} is
\begin{equation}
\Psi_{,\chi\chi}-\tfrac16\Psi_{,\tilde\alpha\tilde\alpha}+\mu\,\ee^{\ell}\Psi=0,\qquad \ell\equiv6\tilde\alpha-\lambda\chi .
\label{eq:wdwexp}
\end{equation}
The exponent $\ell$ has a gradient whose norm with respect to the operator is $\lambda^2-6$, which is spacelike for $\lambda>\sqrt6$, null for $\lambda=\sqrt6$ and timelike for $\lambda<\sqrt6$.

For $\lambda>\sqrt6$ the Wheeler--DeWitt equation can be solved exactly. We define $m=\chi-\lambda\tilde\alpha$, $\ell=6\tilde\alpha-\lambda\chi$ and $z=2\sqrt{\mu/(\lambda^2-6)}\;\ee^{\ell/2}$. With $\partial_\chi=-\lambda\partial_\ell+\partial_m$ and $\partial_{\tilde\alpha}=6\partial_\ell-\lambda\partial_m$ the mixed term cancels and $\Psi_{,\chi\chi}-\tfrac16\Psi_{,\tilde\alpha\tilde\alpha}=(\lambda^2-6)\big(\Psi_{,\ell\ell}-\tfrac16\Psi_{,mm}\big)$. For $\Psi=\ee^{\ii pm}f(\ell)$ the equation becomes $f''+\tfrac{p^2}{6}f+\tfrac{\mu}{\lambda^2-6}\ee^\ell f=0$, and with $f(\ell)=\mathcal C(z)$ and $z\propto\ee^{\ell/2}$ this is $z^2\mathcal C''+z\mathcal C'+(z^2+\tfrac23p^2)\mathcal C=0$, Bessel's equation of order $\ii\sqrt{2/3}\,p$. For every real $p$ the function
\begin{equation}
\Psi_p=\ee^{\ii pm}\,\mathcal C_{\ii\nu}(z),\qquad \nu=p\sqrt{2/3},
\label{eq:Psip}
\end{equation}
is therefore a solution, where $\mathcal C_{\ii\nu}$ is any solution of Bessel's equation of imaginary order $\ii\nu$, for example $J_{\pm\ii\nu}$, $H^{(1)}_{\ii\nu}$ or $H^{(2)}_{\ii\nu}$, and superpositions over $p$ are also solutions.

We verified \eqref{eq:Psip} numerically to $10^{-29}$ for $J_{\ii\nu}$, $H^{(1)}_{\ii\nu}$ and $H^{(2)}_{\ii\nu}$ with arbitrary precision arithmetic. The classical region is $z\gg1$: the Hamilton--Jacobi equation $\frac12G^{AB}S_{,A}S_{,B}+\tilde a^3U=0$ for $S=S(\ell)$ gives $S'=\pm\sqrt{2V_0/(\lambda^2-6)}\,\ee^{\ell/2}$, that is $S=\pm z$, which is the phase of $H^{(1,2)}_{\ii\nu}(z)\sim z^{-1/2}\ee^{\pm\ii z}$. Along the classical scaling solution $m=\text{const}=0$, so it corresponds to $p=0$, and the singularity $T\to0$ lies at $\ell\to+\infty$, because $\ell=\big[(12-2\lambda^2)/\lambda^2\big]\ln T+\text{const}$. The region $z\ll1$, where the wave function is oscillatory in $\ell$ with $\mathcal C\sim z^{\pm\ii\nu}$, corresponds to $\ell\to-\infty$, that is to large Einstein-frame volume at fixed $m$, and there the potential is negligible.

The sector $p=0$ contains the classical ekpyrotic attractor and can be analysed completely. Take $\Psi=H^{(1)}_0(z)+r\,H^{(2)}_0(z)$ with complex $r$ and $|r|\neq1$. For real argument $H^{(1,2)}_0=J_0\pm\ii Y_0$, so that $\Psi=aJ_0+bY_0$ with $a=1+r$ and $b=\ii(1-r)$. Then $\mathrm{Im}(\bar\Psi\Psi')=\mathrm{Im}(\bar ab)(J_0Y_0'-Y_0J_0')$, with $\mathrm{Im}(\bar ab)=\mathrm{Re}[(1+\bar r)(1-r)]=1-|r|^2$ and $J_0Y_0'-Y_0J_0'=2/(\pi z)$, which gives
\begin{equation}
S_{,z}=\frac{(1-|r|^2)\,W}{|\Psi|^2},\qquad W=\frac{2}{\pi z}.
\label{eq:Spz}
\end{equation}
Since $\Psi$ depends on $\ell$ only, $S$ depends on $\ell$ only, and the guidance velocity is proportional to $G^{AB}\partial_B\ell$, which is a fixed direction at constant $m$ with $\dot\ell\propto(\lambda^2-6)S_{,\ell}$. The sign of $S_{,\ell}=S_{,z}\,z/2$ is that of $1-|r|^2$ for all $z$, so no trajectory has a turning point of $\ell$, and every trajectory runs monotonically between $\ell=-\infty$ and the singularity $\ell=+\infty$. For $|r|=1$ the state is real up to a constant phase, $S$ is constant and the Bohmian configuration is static.

We confirmed numerically for $r=0.4$ that $S_{,z}$ agrees with this expression and is positive on $10^{-4}\le z\le10^2$. The relative size of the quantum term at the singularity is
$|\mathcal Q|/|\tfrac12G^{AB}S_{,A}S_{,B}|=O(z^{-2})\to0$ for $z\to\infty$ in this sector, since $\mathcal R\propto z^{-1/2}$, so the Bohmian dynamics approaches the classical crunch. The result is an exact negative statement for the sector that contains the classical ekpyrotic attractor.

States with a non-zero momentum $p$ require a separate treatment. For a single mode $\Psi_p=\ee^{\ii pm}f(\ell)$, with $f$ any solution of $f''+\tfrac{p^2}{6}f+\tfrac{\mu}{\lambda^2-6}\ee^{\ell}f=0$, the real coefficients of this equation give $\frac{\dd}{\dd\ell}\,\mathrm{Im}(\bar ff')=\mathrm{Im}(\bar f f'')=0$, so the flux $j=\mathrm{Im}(\bar ff')$ is independent of $\ell$. Writing $S=pm+\arg f$, one has $S_{,\ell}=j/|f|^2$, and the guidance equations \eqref{eq:guid} give
\begin{equation}
\ell'=(\lambda^2-6)\frac{j}{|f|^2},\qquad \tilde\alpha'=\frac{\lambda p}{6}-\frac{j}{|f|^2},\qquad m'=-\frac{(\lambda^2-6)\,p}{6}.
\label{eq:pflow}
\end{equation}
Hence for $j\neq0$ the coordinate $\ell$ is monotonic along every trajectory, and $\tilde\alpha$ has a critical point exactly where $|f|^2=6j/(\lambda p)$; with $p>0$, $j>0$ one has $\tilde\alpha'>0$ for $|f|^2$ above this level and $\tilde\alpha'<0$ below it, and the case $p<0$ or $j<0$ is obtained by time reversal. For $z\to\infty$, $|f|^2\simeq(\text{const})/z\to0$, so $\tilde\alpha'\to-\infty$: the final stage of every such trajectory is the classical crunch. For $z\ll1$ the solution is $f\simeq a\,z^{\ii\nu}+b\,z^{-\ii\nu}$, which is the free massless solution in the boosted variables; then $|f|^2=|a|^2+|b|^2+2\mathrm{Re}(a\bar b\,z^{2\ii\nu})$ oscillates in $\ell$ and has critical points alternately a maximum and a minimum of $\tilde\alpha$ whenever the level $6j/(\lambda p)=\sqrt6(|a|^2-|b|^2)/\lambda$ lies between $(|a|-|b|)^2$ and $(|a|+|b|)^2$.
Numerically, for $\lambda=3$, $\mu=2$, $p=1$ and $f=aJ_{\ii\nu}+bJ_{-\ii\nu}$: for $b=0$ there is a single critical point, a maximum of $\tilde\alpha$ (a recollapse), at $\ell=-1.37$; for $a=1$, $b=0.95\,\ee^{0.3\ii}$ there are eleven alternating critical points in $-14\le\ell\le4$, so the trajectory bounces several times, the last critical point is a maximum, and the history ends in the crunch; for $b=a$ the flux vanishes and the state is static.
Bounces therefore occur in single-$p$ states only in the weak-potential region $z\lesssim1$, by the same interference mechanism as in Sec.~\ref{sec:massless}, and they do not remove the final crunch at $\ell\to+\infty$. Wave packets in $p$ are not analysed: for them the two travelling waves at large $z$ interfere with different weights, and the estimate of the quantum term above does not apply without further work.

The marginal slope $\lambda=\sqrt6$ is special because the potential then depends on a single null coordinate. For $\lambda=\sqrt6$ the exponent is $6\tilde\alpha-\sqrt6\chi=-\sqrt6x_-$ in the null coordinates $x_\pm=\chi\pm\sqrt6\tilde\alpha$ of Sec.~\ref{sec:massless}, and the equation is $4\Psi_{,x_+x_-}+\mu\,\ee^{-\sqrt6x_-}\Psi=0$.
With $s=\ee^{-\sqrt6x_-}$ it takes the constant-coefficient form
\begin{equation}
\Psi_{,x_+s}=c\,\Psi,\qquad c=\frac{\mu}{4\sqrt6},\qquad s\in(0,\infty).
\label{eq:telegraph}
\end{equation}
This is hyperbolic, and the Goursat problem with data on the characteristics $x_+=\text{const}$ and $s=\text{const}$ has a unique solution, with the Riemann function $I_0\big(2\sqrt{c\,(x_+-x_+')(s-s')}\big)$. The solution space therefore contains two arbitrary functions, as in the massless case. The plane-wave solutions
\begin{equation}
\psi_k=\exp\!\Big[\ii k\,x_+-\ii\,\frac{c}{k}\,s\Big],\qquad k\in\mathbb R\setminus\{0\},
\label{eq:modes}
\end{equation}
span only the part of the solution space that is bounded in $x_+$. The full space is parametrised by two free functions, and modes of both signs of $k$ occur, so interference between counter-propagating components is possible.
The stationary-phase condition gives the classical ray $x_+=-(c/k^2)\,s$, along which
\begin{equation}
\tilde\alpha=\frac{x_+-x_-}{2\sqrt6},\qquad \tilde\alpha_{\max}=-\frac{1}{12}\Big[1+\ln\frac{\mu}{4k^2}\Big],
\label{eq:amax}
\end{equation}
and $\tilde\alpha\to-\infty$ at both ends, a recollapse from and to a singularity with one maximum. For $\mu=2$, $k=1$ this gives $-0.02557$, which agrees with the numerical integration of the guidance equations for the single mode. A single mode has constant modulus, so $\mathcal Q=0$ and the trajectory is the classical ray.
Since $\psi_{-k}=\psi_k^*$, any state with $C(-k)=C(k)^*$ is real up to a constant phase and describes a static Bohmian configuration. For two-mode states $\Psi=A\psi_{k_1}+B\psi_{k_2}$ with $k_1k_2<0$ and unequal weights the flow is non-trivial. In the null coordinates the guidance equations are $x_+'=2S_{,x_-}$, $x_-'=2S_{,x_+}$ and $\tilde\alpha'=(S_{,x_-}-S_{,x_+})/\sqrt6$, so a critical point of $\tilde\alpha$ requires $S_{,x_+}=S_{,x_-}=-\sqrt6\,sS_{,s}$. Numerical integration of these exact equations ($\mu=2$, trajectories started at $(x_+,\ln s)=(-0.2,0.5)$ and $(0.3,-0.5)$ and integrated in both directions) for $(k_1,k_2,B/A)=(1,-1,0.5)$, $(1,-1,0.9\,\ee^{0.7\ii})$ and $(1,-1.5,0.7\,\ee^{0.9\ii})$ gives trajectories that are monotonic or have a single maximum of $\tilde\alpha$ and run to $\tilde\alpha\to-\infty$ at both ends, apart from critical-point pairs of amplitude below $10^{-3}$ that we regard as unresolved. For $(1,-0.6,0.8\,\ee^{0.4\ii})$ the numerical $\tilde\alpha$ oscillated rapidly and we did not resolve it. These states are plane waves and are not normalisable, so the results are statements about the flow of these specific states and not a theorem.

Finally the classical contraction can be transcribed to the Jordan frame. The Jordan-frame scale factor is $a_J=\tilde a/\sqrt F$, with $\dd t_J=\dd\tilde t/\sqrt F$. In the scaling solution of Sec.~\ref{sec:exp}, $\chi\to-\infty$ as $T\to0$. For finite $\kappa$ and $F=1+\xi\phi^2$, Eq.~\eqref{eq:slopes}
gives $\ln F\simeq2\sqrt\xi\,|\chi|=-(4\sqrt\xi/\lambda)\ln T$, so $a_J\propto T^{2/\lambda^2+2\sqrt\xi/\lambda}$ and $t_J\propto T^{1+2\sqrt\xi/\lambda}$. In the Jordan frame the contraction is therefore a power law with
\begin{equation}
\epsilon_J=\frac{\lambda^2+2\lambda\sqrt\xi}{2+2\lambda\sqrt\xi}\quad(\text{finite }\kappa),\qquad
\epsilon_J^{\rm metric}=\frac{\lambda^2+2\lambda s}{2+2\lambda s},\quad s=\sqrt{\frac{\xi}{1+6\xi}}\quad(\kappa\to\infty).
\label{eq:epsJ}
\end{equation}
For $\xi=0$ both reduce to $\lambda^2/2$. As $\xi\to\infty$, $\epsilon_J\to1$ for finite $\kappa$. The Jordan-frame contraction is ekpyrotic, $\epsilon_J>3$, if and only if $\lambda^2-4\lambda\sqrt\xi-6>0$, that is $\sqrt\xi<(\lambda^2-6)/(4\lambda)$
for finite $\kappa$, and the same condition with $s$ in place of $\sqrt\xi$ for the torsion-free theory. For a given $\lambda$ the allowed range of $\xi$ is therefore larger without torsion, because $s<\sqrt\xi$. Torsion thus makes the Jordan-frame contraction less ekpyrotic at strong coupling.

echo ok

\section{Perturbations and inflation}
\label{sec:pert}

The perturbation theory used for both inflation and ekpyrotic contraction is classical, and it is the description that applies in the semiclassical regime $|\mathcal Q|\ll|\tfrac12G^{AB}S_{,A}S_{,B}|$ of Sec.~\ref{sec:formalism}.
It is derived here for the reduced theory \eqref{eq:Seff}. The algebraic elimination of the torsion trace holds for arbitrary inhomogeneous configurations: the covariant field equation for $T_\mu$ gives $T_\mu=-\tfrac32F'\partial_\mu\phi/(F+\kappa)$, the axial and tensor parts vanish if their coefficients are non-degenerate,
and the effective action is \eqref{eq:Seff}. The Weyl rescaling $\tilde g_{\mu\nu}=F\,g_{\mu\nu}$ with the field redefinition $\dd\chi=\sqrt K\,\dd\phi$ then gives, exactly,
\begin{equation}
S_E=\int\dd^4x\sqrt{-\tilde g}\left[\tfrac12\tilde R-\tfrac12\tilde g^{\mu\nu}\partial_\mu\chi\,\partial_\nu\chi-U(\chi)\right],\qquad U=\frac V{F^2},
\label{eq:SE}
\end{equation}
which is the covariant version of Eq.~\eqref{eq:Leinstein}. It is a canonical single-field theory, so its linear perturbations are those of a minimally coupled scalar field, and the torsion dependence is in the function $\chi(\phi)$ only.

For the scalar sector, in the comoving gauge $\delta\chi=0$ the curvature perturbation $\zeta$ has the quadratic action $\tfrac12\int\dd\tilde\tau\,\dd^3x\,\tilde z^2[(\zeta')^2-(\nabla\zeta)^2]$ with $\tilde z=\tilde a\sqrt{2\epsilon}$, $\epsilon=-\dot{\tilde H}/\tilde H^2$ and $\tilde\tau$ the Einstein-frame conformal time.
The Mukhanov--Sasaki variable $v=\tilde z\zeta$ obeys
\begin{equation}
v_k''+\Big(k^2-\frac{\tilde z''}{\tilde z}\Big)v_k=0 .
\label{eq:MS}
\end{equation}
For the tensor sector, each polarisation has a canonically normalised variable $v_T\propto\tilde a\,h$ obeying $v_{T,k}''+(k^2-\tilde a''/\tilde a)v_{T,k}=0$, with $\mathcal P_T=(8/2\pi^2)\,k^3|v_T/\tilde a|^2$ and $\mathcal P_\zeta=(k^3/2\pi^2)|v/\tilde z|^2$ for modes normalised to the Bunch--Davies vacuum $v_k\to\ee^{-\ii k\tilde\tau}/\sqrt{2k}$ at early times.
These normalisations reproduce the standard inflationary results $\mathcal P_T=2H^2/\pi^2$ and $\mathcal P_\zeta=H^2/(8\pi^2\epsilon)$.

The scalar spectrum is also independent of the frame. The comoving gauge $\delta\chi=0$ is the gauge $\delta\phi=0$, on which $F$ is uniform. The Weyl factor is then a spatially constant function of time on these slices, and the curvature perturbation of the spatial metric is unchanged by it. Hence $\zeta$ is the same in the Jordan and Einstein frames,
and likewise the tensor perturbation $h_{ij}$ is unchanged. Spectra of $\zeta$ and $h_{ij}$ are therefore frame independent, while the background quantities $\epsilon$, $a$ and $H$ differ between frames and are to be quoted with the frame stated.

For power-law backgrounds the spectra follow in closed form. For constant $\epsilon$ in the Einstein frame the conformal-time scale factor is $\tilde a\propto(-\tilde\tau)^p$ with $p=1/(\epsilon-1)$ for $\epsilon\neq1$ and $\tilde\tau<0$ (contraction for $\epsilon>1$ with $T\to0$, inflation for $\epsilon<1$). Since $\tilde z\propto\tilde a$ for constant $\epsilon$, both \eqref{eq:MS} and the tensor equation have
$\tilde z''/\tilde z=\tilde a''/\tilde a=p(p-1)/\tilde\tau^2\equiv(\nu^2-\tfrac14)/\tilde\tau^2$, so $\nu=|p-\tfrac12|$ and the solution that is normalised to the vacuum is $v_k=\tfrac12\sqrt{\pi(-\tilde\tau)}\,H^{(1)}_\nu(-k\tilde\tau)$ up to a phase.
For $\nu>0$ the small-argument behaviour $H^{(1)}_\nu(x)\propto x^{-\nu}$ gives, outside the horizon, $|v_k|^2\propto k^{-2\nu}$ and hence
\begin{equation}
\mathcal P_\zeta\propto\mathcal P_T\propto k^{3-2\nu},\qquad n_s-1=n_t=3-2\nu,\qquad r=16\epsilon .
\label{eq:powerlaw}
\end{equation}
The equality of the tilts and $r=16\epsilon$ hold because $v$ and $v_T$ obey the same equation, and $r=\mathcal P_T/\mathcal P_\zeta=8\cdot2\epsilon$ follows from the normalisations above.
The relation \eqref{eq:powerlaw} holds when $\epsilon$ varies slowly, $|\dd\ln\epsilon/\dd\ln\tilde a|\ll1$; otherwise $\nu$ acquires corrections of the order of that rate, and the expressions below are leading-order estimates.

\begin{figure}[t]
\centering
\includegraphics[width=\textwidth]{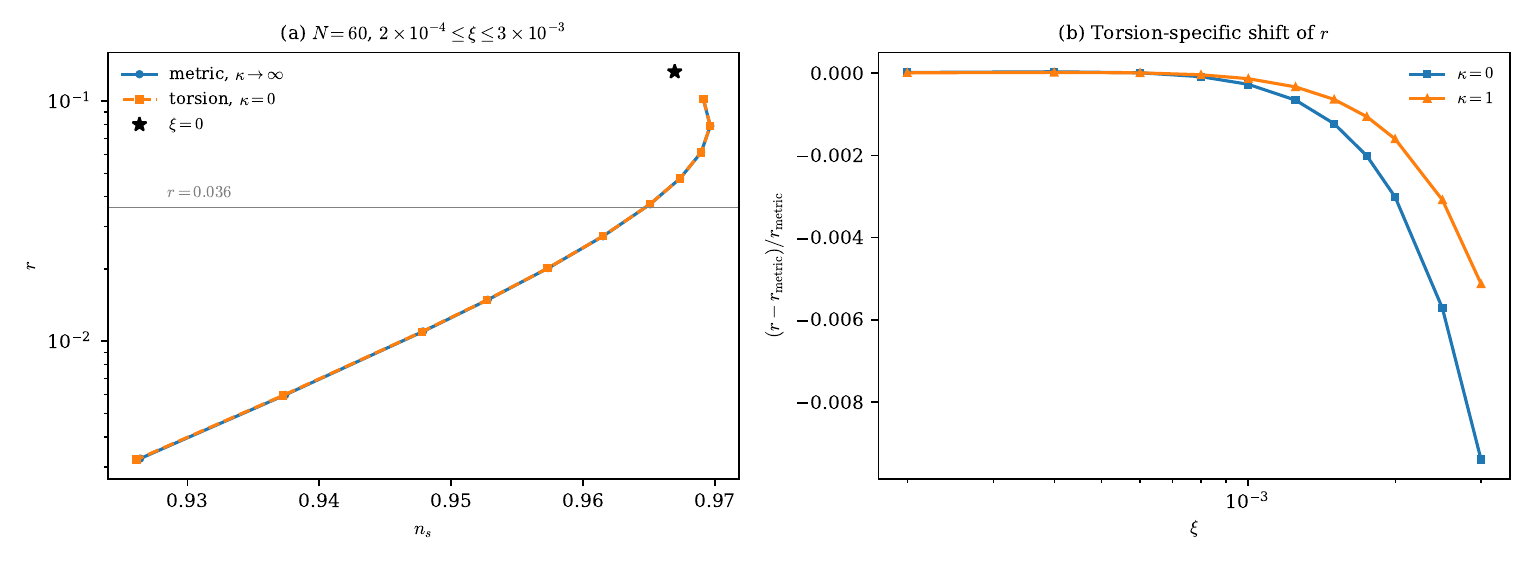}
\caption{(a) Tensor-to-scalar ratio against spectral index for $2\times10^{-4}\le\xi\le3\times10^{-3}$ and $N=60$; the two curves are
nearly indistinguishable. The star marks $\xi=0$. (b) Relative difference of $r$ from the torsion-free value for $\kappa=0$ and $\kappa=1$.}
\label{fig:infl}
\end{figure}

We now apply this framework to chaotic inflation with $V=\tfrac12m^2\phi^2$ and $F=1+\xi\phi^2$, using the exact Einstein-frame problem rather than an expansion in $\xi$.
The Einstein-frame potential is $U=m^2\phi^2/[2(1+\xi\phi^2)^2]$, which has a maximum at $\phi=1/\sqrt{3\xi}$. Inflation takes place on the
side $\phi<1/\sqrt{3\xi}$, so only $\xi\lesssim10^{-2}$ is relevant. We use the slow-roll expressions
\begin{equation}
\epsilon=\frac{1}{2K}\Big(\frac{U_{,\phi}}{U}\Big)^2,\qquad \eta=\frac{U_{,\chi\chi}}{U},\qquad
N=\int_{\phi_{\rm end}}^{\phi_*}\frac{K\,U}{U_{,\phi}}\,\dd\phi,\qquad r=16\epsilon,\quad n_s=1-6\epsilon+2\eta,
\end{equation}
with $\epsilon(\phi_{\rm end})=1$ and $N=60$. The integrals are evaluated numerically without expansion in $\xi$. Table~\ref{tab:infl} and Fig.~\ref{fig:infl}
give the results for $\kappa=0$ and $\kappa\to\infty$. The overall amplitude fixes $m$ and does not affect $r$ and $n_s$.

\begin{table}[t]
\centering
\caption{Slow-roll observables for $N=60$ with $F=1+\xi\phi^2$ and $V=\tfrac12m^2\phi^2$.}
\label{tab:infl}
\begin{tabular}{lcccc}
\toprule
 & \multicolumn{2}{c}{torsion-free ($\kappa\to\infty$)} & \multicolumn{2}{c}{Einstein--Cartan ($\kappa=0$)}\\
$\xi$ & $r$ & $n_s$ & $r$ & $n_s$\\
\midrule
0 & 0.1322 & 0.9669 & 0.1322 & 0.9669\\
$1.0\times10^{-3}$ & 0.03717 & 0.96507 & 0.03716 & 0.96506\\
$1.5\times10^{-3}$ & 0.02016 & 0.95735 & 0.02014 & 0.95731\\
$2.0\times10^{-3}$ & 0.01098 & 0.94786 & 0.01095 & 0.94777\\
$3.0\times10^{-3}$ & 0.00325 & 0.92641 & 0.00322 & 0.92615\\
\bottomrule
\end{tabular}
\end{table}

The non-minimal coupling reduces $r$ strongly: $r$ falls from $0.132$ at $\xi=0$ to $0.037$ at $\xi=10^{-3}$, with $n_s$ close to $0.965$. This effect
is already present in the torsion-free theory and is the known effect of the coupling \cite{Futamase1989,Fakir1990}. The torsion-specific shift, the difference between
$\kappa=0$ and $\kappa\to\infty$, is at most $0.3\times10^{-3}$ of $r$ at $\xi=10^{-3}$ and below $1\%$ of $r$ at $\xi=3\times10^{-3}$, and
$|\Delta n_s|<3\times10^{-4}$ throughout. This is consistent with the order estimate of Sec.~\ref{sec:formalism}, since $K-K_{\rm metric}=O(\xi^2)$ while the coupling effect is $O(\xi)$.
For the comparison with data we use the bound $r<0.036$ at 95\% confidence \cite{BK18} and $n_s=0.9649\pm0.0042$ \cite{Planck2018}. At $N=60$ the bound on $r$ requires $\xi\gtrsim1.03\times10^{-3}$, and $n_s$ stays within two standard deviations of the Planck value for
$\xi\lesssim1.5\times10^{-3}$. The conclusion, that the coupling can bring the quadratic model into the allowed region in a narrow window and that torsion does not change this, depends on $N$,
on the post-inflationary history and on the data set. More recent combinations of ground-based and satellite data prefer somewhat larger $n_s$, and the numbers in Table~\ref{tab:infl} allow the comparison to be redone.
No bispectrum is computed here. The reduced theory is a canonical single-field model with a slow-roll attractor, for which the standard consistency relation for the squeezed limit applies
if the perturbations start in the adiabatic vacuum; a primordial state modified by a preceding bounce would require a separate calculation.

\section{Ekpyrotic contraction and comparison with inflation}
\label{sec:ekpyrosis}

The alternative to inflation considered here is a phase of slow contraction with $w>1$, equivalently $\epsilon>3$, driven by the same scalar field and the same torsionful gravity. Sections~\ref{sec:exp}--\ref{sec:exp} supply the background. This section gives the perturbations in the contracting phase.

We turn to ekpyrotic contraction and begin with the perturbations in the scaling solution of Sec.~\ref{sec:exp}. There $\epsilon=\lambda^2/2$ is constant and $\tilde a\propto(-\tilde\tau)^p$ with $p=1/(\epsilon-1)$, as derived in Sec.~\ref{sec:exp}. Then $\nu=|p-\tfrac12|=(\epsilon-3)/[2(\epsilon-1)]$ for $\epsilon>3$, and \eqref{eq:powerlaw} gives
\begin{equation}
n_s-1=n_t=\frac{2\epsilon}{\epsilon-1},\qquad r=16\epsilon,
\label{eq:ekpspectra}
\end{equation}
so that $n_s=3+2/(\epsilon-1)$, between $3$ and $4$ for $\epsilon>3$. With $\epsilon>3$ one also has $r>48$. The curvature perturbation is constant outside the horizon in this phase, since the exponent of $(-\tilde\tau)$ in $\zeta=v/\tilde z$ vanishes for $\nu=\tfrac12-p$.

The result holds for every $\kappa$ and every $\xi$ in the family of Sec.~\ref{sec:exp}, because $\epsilon=\lambda^2/2$ is a property of the Einstein-frame potential: in this family the spectra of a single ekpyrotic field are independent of torsion and of the non-minimal coupling.
Since the spectrum is the same for the torsion-free and the torsionful theory, torsion is not a way to repair the blue tilt. The observed spectrum is nearly scale invariant and has $r<0.036$ \cite{BK18,Planck2018}, so a single-field ekpyrotic phase of this kind cannot produce the observed perturbations by itself, whatever the value of the torsion constant.
The perturbations of the contracting phase would also have to be matched through the bounce or crunch, which depends on the model of the transition and is not considered here.

Torsion matters for the ekpyrotic background in a different way. A Jordan-frame exponential potential $V=-V_0\ee^{-\beta\phi}$ with $F=1+\xi\phi^2$ gives an Einstein-frame potential $U=-V_0\ee^{-\beta\phi}/F^2$ with $\epsilon(\phi)=(2K)^{-1}(\beta+2F'/F)^2$. For $\xi=0$ this is the constant $\beta^2/2$. For $\xi\neq0$ and $\phi\to-\infty$, where the potential becomes more negative, $\epsilon\to\infty$ in both cases, because $K\simeq1/(\xi\phi^2)$ for finite $\kappa$ and $K\simeq(1+6\xi)/(\xi\phi^2)$ for $\kappa\to\infty$:
\begin{equation}
\frac{\epsilon_{\rm torsion}}{\epsilon_{\rm metric}}\;\xrightarrow[\phi\to-\infty]{}\;1+6\xi .
\label{eq:epsratio}
\end{equation}
Numerically, for $\beta=2$ and $\xi=0.1$ the ratio is $1.28$, $1.55$, $1.59$, $1.60$ at $\phi=-3,-10,-30,-100$, approaching $1.6=1+6\xi$, and the values of $\epsilon_{\rm torsion}$ are $1.8$, $15$, $159$, $1923$. The non-minimal coupling therefore turns a potential that is not steep enough at $\xi=0$ ($\epsilon=2$) into an ekpyrotic one at large field, and torsion makes it steeper by the factor $1+6\xi$.
This is a torsion-specific effect on the background, and it is of order one for $\xi\gtrsim0.1$. However, $\epsilon$ then grows quickly along the contraction. With $\mathrm{d}\chi/\mathrm{d}\ln\tilde a=+\sqrt{2\epsilon}$ (both $\chi$ and $\ln\tilde a$ decrease) and $\phi\propto\ee^{\sqrt\xi|\chi|}$ one has $\mathrm{d}\ln\epsilon/\mathrm{d}\ln\tilde a\sim2\sqrt{2\xi\epsilon}$, which is not small; for example it is $\simeq2.8$ for $\xi=0.1$ and $\epsilon=10$. Then \eqref{eq:ekpspectra} is not valid and a scaling-solution estimate of the spectrum is not available.
The definite statements of this section are made for the Einstein-frame exponential family, where $\epsilon$ is exactly constant, and Sec.~\ref{sec:ekpyrosis} tests the running case numerically.

The leading-order tilt \eqref{eq:ekpspectra} assumes constant $\epsilon$. To test it we integrated the Mukhanov--Sasaki equation \eqref{eq:MS} on exact backgrounds of the reduced theory. With $x=-\ln\tilde a$ and Bunch--Davies data imposed at $k/(\tilde a\tilde H)=200$, the equation for $\zeta$ reads $\zeta_{,xx}+(\mathrm d\ln|\tilde H|/\mathrm dx-3+\mathrm d\ln\epsilon/\mathrm dx)\zeta_{,x}+k^2\ee^{2x}\tilde H^{-2}\zeta=0$. For the exact scaling background with $\epsilon=4.5$ the numerical spectrum has tilt $2.571$ against the analytic $2\epsilon/(\epsilon-1)=2.5714$, and its amplitude agrees with the Hankel solution to $0.44\%$, which validates the numerics.
For the Jordan-frame exponential $V=-V_0\ee^{-2\phi}$ with $F=1+0.05\,\phi^2$ and $\kappa=0$, started on the instantaneous attractor with $\epsilon=4$, the value of $\epsilon$ at horizon exit of the six wavenumbers lies between $15.7$ and $22.7$ and the numerical tilt is $n_s-1=1.88$ to $1.91$, whereas \eqref{eq:ekpspectra} evaluated with the local $\epsilon$ gives $2.10$ to $2.13$. The running of $\epsilon$ ($\mathrm d\ln\epsilon/\mathrm d\ln\tilde a\simeq2.8$) lowers the tilt by about $0.2$, but the spectrum remains strongly blue ($n_s\simeq2.9$). The conclusion that a single ekpyrotic field is excluded is therefore not an artefact of the constant-$\epsilon$ approximation. The corresponding torsion-free integration was not carried out.

The failure of the single-field scenario is usually repaired by an entropic mechanism. The standard way to obtain a scale-invariant spectrum from an ekpyrotic phase is the conversion of the entropy perturbation of a second field into $\zeta$ \cite{Lehners2007,Buchbinder2007,Creminelli2007,Koyama2007,Lehners2008}. We treat it in the present setting for a second scalar $\sigma$ with Jordan-frame action $-\tfrac12(\partial\sigma)^2$, no coupling to $F$, and a joint potential $V(\phi,\sigma)$. The Weyl rescaling of Sec.~\ref{sec:pert} gives
\begin{equation}
S_E=\int\dd^4x\sqrt{-\tilde g}\left[\tfrac12\tilde R-\tfrac12\,\dd\chi^2-\frac{1}{2F}(\partial\sigma)^2-U(\chi,\sigma)\right],\qquad U=\frac{V}{F^2},
\label{eq:twofield}
\end{equation}
that is, a sigma model with field-space metric $\dd\chi^2+F^{-1}\dd\sigma^2$. Its Gaussian curvature is $K_{\rm fs}=-(F^{-1/2})_{,\chi\chi}F^{1/2}=\tfrac12s'-\tfrac14s^2$ with $s=F_{,\chi}/F$. In the asymptotic regime $\chi\to-\infty$ of the ekpyrotic roll, $F\propto\ee^{-s_0\chi}$ with constant $s_0$, so $K_{\rm fs}\to-s_0^2/4$, where from Eq.~\eqref{eq:slopes}
\begin{equation}
s_0=2\sqrt\xi\quad(\text{finite }\kappa),\qquad s_0=2\sqrt{\xi/(1+6\xi)}\quad(\kappa\to\infty).
\end{equation}
Torsion therefore makes the field space asymptotically hyperbolic with curvature $-\xi$, whereas the torsion-free value is $-\xi/(1+6\xi)$. We choose
\begin{equation}
U(\chi,\sigma)=-U_0\,\ee^{-\lambda\chi}\Big[1+\frac{g\,\sigma^2}{2F(\chi)}\Big],\qquad g>0,
\end{equation}
which has the scaling solution of Sec.~\ref{sec:exp} along the geodesic $\sigma=0$ (the metric does not depend on $\sigma$ and $U_{,\sigma}=0$ there), with $\epsilon=\lambda^2/2$ and zero turn rate. The $1/F$ in the bracket keeps the entropic mass proportional to $\tilde H^2$ and is part of the definition of the model; in Jordan variables it is $V(\phi,\sigma)=-V_0F^2\ee^{-\lambda\chi(\phi)}[1+g\sigma^2/(2F)]$.

For zero turn rate the entropy perturbation $\delta s=F^{-1/2}\delta\sigma$ decouples from $\zeta$ \cite{Gordon2001,Peterson2011}, and $u=\tilde a\,\delta s$ obeys $u''+(k^2-\tilde a''/\tilde a+\tilde a^2m_s^2)u=0$ with $m_s^2=V_{NN}+\epsilon\tilde H^2R_{\rm fs}$, where $R_{\rm fs}=2K_{\rm fs}$ and $V_{NN}$ is the covariant second derivative of $U$ along the unit normal. With $\Gamma^\chi_{\sigma\sigma}=F_{,\chi}/(2F^2)$ this is $V_{NN}=F\,U_{,\sigma\sigma}-\tfrac12 s\,U_{,\chi}$. On the scaling solution $|U|=(\epsilon-3)\tilde H^2$ and $U_{,\chi}=\lambda|U|$, so
\begin{equation}
\frac{m_s^2}{\tilde H^2}\equiv M^2=-g(\epsilon-3)-\frac{\lambda s}{2}(\epsilon-3)+\epsilon\Big(s'-\frac{s^2}{2}\Big)\;\longrightarrow\;-g(\epsilon-3)+\frac{\lambda s_0}{2}(\epsilon-3)-\frac{\epsilon s_0^2}{2}.
\label{eq:M2}
\end{equation}
We tested \eqref{eq:M2} against a direct nonlinear integration of the coupled equations for $\tilde a$, $\chi$ and $\sigma$ with $F=\ee^{-s_0\chi}$ exactly, $\lambda=3$ ($\epsilon=4.5$), $s_0=0.4$ and $g=2$. For $k=0$ the proper entropy field scales as $T^q$ with $q(q-1)+3q/\epsilon+M^2/\epsilon^2=0$, which gives $q=-0.2197$. The integrated $\delta s$ approaches this: the fitted exponent is $-0.143$, $-0.185$, $-0.2095$ and $-0.2165$ in the windows $T\in(0.05,0.1)$, $(0.01,0.05)$, $(0.002,0.01)$ and $(0.0005,0.002)$, converging with the expected power-law decay of the subdominant mode.

With $\tilde a\propto(-\tilde\tau)^p$, $p=1/(\epsilon-1)$ and $\tilde a^2m_s^2=M^2p^2/\tilde\tau^2$, the Bunch--Davies solution has index
\begin{equation}
\nu_s^2=\Big(p-\frac12\Big)^2-M^2p^2,\qquad n_s-1=3-2\nu_s .
\label{eq:nus}
\end{equation}
A later turn of the trajectory converts $\delta s$ into $\zeta$ with a transfer that does not depend on $k$ for super-horizon modes, so the tilt of $\zeta$ is that of $\delta s$. The conversion efficiency, the amplitude and the non-Gaussianity are not computed here.
For $s_0=0$ the spectrum is exactly scale invariant ($\nu_s=3/2$) if $M^2=-\epsilon(2\epsilon-3)$, that is $g_0=\epsilon(2\epsilon-3)/(\epsilon-3)$. For $s_0\neq0$ at fixed $g=g_0$, to first order in $s_0$,
\begin{equation}
\delta n_s\simeq\frac{\lambda s_0(\epsilon-3)}{3(\epsilon-1)^2}\;\propto\;\sqrt\xi,
\end{equation}
which is non-analytic in $\xi$ and arises from the connection term $-\tfrac12sU_{,\chi}$, that is from the dependence of the normalisation of the entropy field on $F$. The curvature term $\epsilon R_{\rm fs}$ contributes at order $\xi$. The torsion-specific part of the tilt is the difference between $s_0=2\sqrt\xi$ and $s_0=2\sqrt{\xi/(1+6\xi)}$ and is of order $\xi^{3/2}$. Table~\ref{tab:entropic} gives numbers from \eqref{eq:M2} and \eqref{eq:nus}.
\begin{table}[t]
\centering
\caption{Spectral index of the entropy mode at fixed $g=g_0$ (scale invariant for $\xi=0$), and the value $g_*$ that restores $n_s=1$.}
\label{tab:entropic}
\begin{tabular}{ccccccc}
\toprule
$\epsilon$ & $\xi$ & $n_s$ (torsion) & $n_s$ (torsion-free) & difference & $g_*$ (torsion) & $g_*$ (torsion-free)\\
\midrule
10 & $10^{-3}$ & 1.00799 & 1.00797 & $2.3\times10^{-5}$ & 24.424 & 24.424\\
10 & $10^{-2}$ & 1.02422 & 1.02357 & $6.5\times10^{-4}$ & 24.704 & 24.693\\
10 & $10^{-1}$ & 1.06574 & 1.05462 & $1.1\times10^{-2}$ & 25.414 & 25.225\\
30 & $10^{-3}$ & 1.00520 & 1.00518 & $1.5\times10^{-5}$ & 63.576 & 63.575\\
30 & $10^{-2}$ & 1.01615 & 1.01570 & $4.5\times10^{-4}$ & 64.086 & 64.065\\
30 & $10^{-1}$ & 1.04806 & 1.03872 & $9.3\times10^{-3}$ & 65.561 & 65.131\\
\bottomrule
\end{tabular}
\end{table}
Unlike the single-field case, torsion therefore does change the entropic tilt, but only at the level of $10^{-3}$ for $\xi\le10^{-2}$; the non-minimal coupling itself shifts the tilt by $0.008$ ($\epsilon=10$) at $\xi=10^{-3}$, which is about twice the Planck uncertainty on $n_s$ \cite{Planck2018} unless $g$ is retuned, and the required retuning of $g$ is $0.6\%$ at $\xi=10^{-3}$ and about $1$ to $2\%$ at $\xi=10^{-2}$. The result holds in the regime where $s$ is constant, that is $\sqrt\xi|\chi|\gg1$, and for the Einstein-frame exponential family.

The quantum aspects of the ekpyrotic phase can be summarised as follows. Section~\ref{sec:exp} gives the exact Wheeler--DeWitt solutions for $\lambda>\sqrt6$, and in the sector that contains the classical ekpyrotic attractor there is no Bohmian bounce (Eq.~\eqref{eq:Spz}). For single-$p$ states bounces occur only in the weak-potential region and the history ends in the crunch, and the quantum term is suppressed at the singularity. The exponential ekpyrotic potential alone therefore does not resolve the singularity in the states studied, in contrast with the massless sector of Sec.~\ref{sec:massless}, where the two-packet interference produces a turning point for every trajectory sampled. The transfer of tensor perturbations through the massless Bohmian bounce is computed in Sec.~\ref{sec:massless}.

\begin{figure}[t]
\centering
\includegraphics[width=\textwidth]{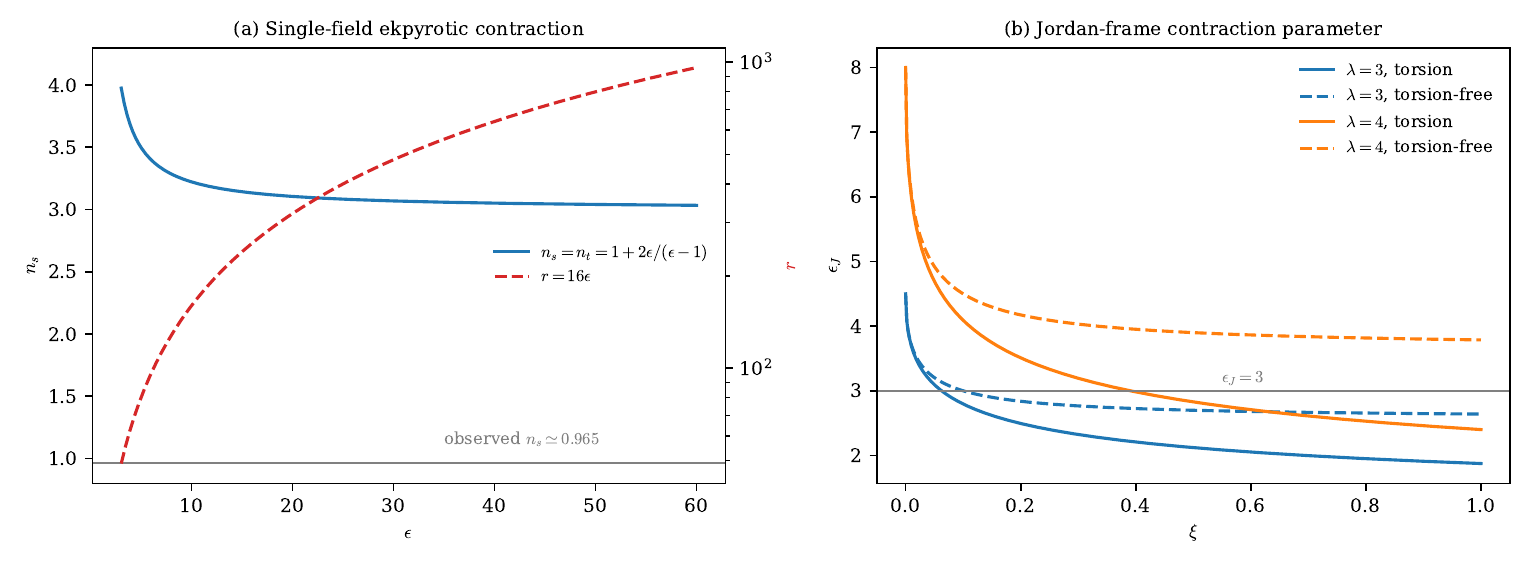}
\caption{(a) Single-field ekpyrotic contraction: spectral index $n_s=n_t$ (left axis) and tensor-to-scalar ratio $r=16\epsilon$ (right axis) against $\epsilon>3$. (b) Jordan-frame contraction parameter $\epsilon_J$ of \eqref{eq:epsJ} against $\xi$ for $\lambda=3$ and $4$, for finite $\kappa$ (solid) and for the torsion-free theory (dashed). The line $\epsilon_J=3$ marks the end of the ekpyrotic regime.}
\label{fig:ekpyrosis}
\end{figure}

The two scenarios can now be compared directly. Both are realised in the same reduced theory \eqref{eq:SE}, and they differ in the sign of $\epsilon-1$ and in the potential. Table~\ref{tab:compare} compares them.

\begin{table}[t]
\centering
\caption{Inflation (quadratic potential, $N=60$) and single-field ekpyrotic contraction in the reduced theory.}
\label{tab:compare}
\begin{tabular}{p{0.28\textwidth}p{0.33\textwidth}p{0.33\textwidth}}
\toprule
 & Inflation & Ekpyrotic contraction\\
\midrule
Potential in $\chi$ & $U=V/F^2>0$, hilltop on one side & $U=-V_0\ee^{-\lambda\chi}$, $\lambda>\sqrt6$\\
Einstein-frame $\epsilon$ & $\ll1$ & $\lambda^2/2>3$\\
Spectral index & $0.965$ at $\xi=10^{-3}$ & $n_s=3+2/(\epsilon-1)\in(3,4)$\\
Tensor ratio $r$ & $0.037$ at $\xi=10^{-3}$ & $r=16\epsilon>48$\\
Tensor tilt & $n_t=-2\epsilon$ & $n_t=n_s-1>2$\\
Torsion-specific effect on spectra & $<1\%$ in $r$, $<3\times10^{-4}$ in $n_s$ ($\xi\le3\times10^{-3}$) & none in the scaling family\\
Torsion-specific effect on background & $K-K_{\rm metric}=O(\xi^2)$ & $\epsilon\to(1+6\xi)\epsilon_{\rm metric}$ at large field; $\epsilon_J$ of \eqref{eq:epsJ}\\
Quantum sector & semiclassical, $\mathcal Q\ll$ classical terms & no Bohmian bounce for $p=0$; crunch classically and quantum mechanically\\
Entropic (two-field) version & not needed & $n_s-1\propto\sqrt\xi$ shift, torsion part $\sim\xi^{3/2}$ (Table~\ref{tab:entropic})\\
Status as sole source of perturbations & consistent with data for $1.03\times10^{-3}\lesssim\xi\lesssim1.5\times10^{-3}$ & single field excluded (blue, $r>1$); needs the entropic mechanism\\
\bottomrule
\end{tabular}
\end{table}

In the common Einstein-frame description the two scenarios correspond to the two signs of $\epsilon-1$, and torsion is encoded in the same function $K(\phi)$ in both. The direct effect of torsion on the spectra is small for inflation and absent in the exact ekpyrotic scaling family. For the ekpyrotic background, however, torsion changes the dynamics at order one once $\xi\phi^2\gtrsim1$, through the factor $1+6\xi$ in $\epsilon$ for a Jordan-frame exponential and through $\epsilon_J$ of \eqref{eq:epsJ} for the Einstein-frame exponential, where the Jordan-frame contraction ceases to be ekpyrotic once $\sqrt\xi>(\lambda^2-6)/(4\lambda)$. The same strong-field regime governs the late-time exponent $1-\sqrt{6\xi}$ of the massless solutions (Eq.~\eqref{eq:slopes}), so this regime is where torsion matters in every sector examined. A single ekpyrotic field is excluded as the source of the primordial spectrum irrespective of torsion, also when the running of $\epsilon$ is included. The entropic mechanism restores scale invariance, and in this setting torsion enters through the curvature and the connection of the hyperbolic field space, shifting the tilt at order $\sqrt\xi$ through the non-minimal coupling and at order $\xi^{3/2}$ through torsion. The quantum sectors also differ. For the massless field, interference produces Bohmian bounces in all trajectories sampled, whereas for the exponential ekpyrotic potential the scaling sector has no bounce, and single-momentum states bounce only in the weak-potential region and then end in the crunch.

\section{Discussion}
\label{sec:discussion}

The central result is structural. After the torsion trace is eliminated, the homogeneous sector of the Riemann--Cartan scalar--tensor model is a minimally coupled scalar field in general relativity, and all dependence on torsion sits in the Einstein-frame kinetic function $K(\phi)$, whose torsion-specific part $K-K_{\rm metric}=-3F'^2/[2F(F+\kappa)]$ is known exactly. This has three consequences that go beyond the individual calculations. The curvature of the minisuperspace metric, which could be mistaken for a signature of torsion, is a conformal-factor effect: it is present without torsion and vanishes for non-constant couplings of the form \eqref{eq:flatODE}. The Laplace--Beltrami Wheeler--DeWitt equation, its quantum potential and its Bohmian flow lines are covariant under the change of frame and of lapse, which turns the question of how torsion affects the quantum dynamics into a question about the map between the Einstein-frame and Jordan-frame variables. And, because the Einstein-frame dynamics is that of general relativity with a scalar field, the Hubble rate obeys $\dot{\tilde H}=-\dot\chi^2/2\le0$ at the classical level, so that no classical bounce of the Einstein-frame scale factor exists for any $F$ and $\kappa$. Torsion sourced by the gradient of a scalar field does not provide the repulsion that spin provides in the Einstein--Cartan bounces of spin fluids and Dirac fermions, and any classical minimum of the Jordan-frame scale factor in this model is an effect of the factor $F^{-1/2}$.

The quantum bounce of the massless scalar illustrates how this works. The interference of the two packets gives a turning point in all the trajectories we integrated, with the exact node structure, the guidance momentum on the symmetry line and the criterion for a minimum of the scale factor given in Sec.~\ref{sec:massless}. The bounce is the same curve in the Einstein-frame configuration space for every $F$ and $\kappa$, so it is a property of the state and not of torsion. Torsion and the non-minimal coupling determine how the curve appears in the Jordan frame. At the symmetric event the depth of the Jordan-frame bounce is changed by $\xi$ alone, through $-\xi p^2$ in the second derivative, and a threshold $\xi_c$ separates a single bounce from two bounces with an intermediate expansion; for the packets studied $\xi_c$ lies between $0.08$ and $0.16$. Torsion enters at order one only at large field, where $\xi\phi^2\gtrsim1$ and the late-time Jordan-frame exponent changes from $1-\sqrt{6\xi/(1+6\xi)}$ to $1-\sqrt{6\xi}$, so that for $\xi>1/6$ the Jordan-frame scale factor decreases at late times while the Einstein-frame one grows. The consequence for the interpretation of bounces in scalar--tensor quantum cosmology is that a statement of singularity resolution is incomplete unless the frame is specified, in line with the clock dependence found for other quantisations \cite{GielenMP2020}, and that the Jordan-frame statement can be changed, qualitatively, by the strength of the coupling although the underlying Bohmian dynamics is not. Tensor modes cross this bounce with transfer coefficients that differ from unity by less than $0.1\%$ for $k=2\times10^{-4}$, so the blue spectrum of the incoming stiff-fluid contraction is preserved, which is the main obstacle for using this sector as a source of primordial perturbations.

The exponential potentials show a contrast. In the sector that contains the classical ekpyrotic attractor the Bohmian flow has a velocity of fixed sign and every trajectory ends in the singularity, and single-momentum states can bounce only in the region where the potential is negligible, again by interference, before collapsing. The steep exponential alone therefore does not resolve the singularity in the states examined, and a bouncing ekpyrotic cosmology in this framework requires additional ingredients at the singularity. At the classical level the comparison with inflation is simple because both scenarios are the two signs of $\epsilon-1$ in the same Einstein-frame theory. For quadratic inflation the model matches the Planck and BICEP/Keck results only for $1.03\times10^{-3}\lesssim\xi\lesssim1.5\times10^{-3}$ at $N=60$, and the torsion-specific shift of $r$ is below one per cent, so these data do not distinguish torsion from the metric theory in this model. In the ekpyrotic case a single field gives $n_s=3+2/(\epsilon-1)$ and $r=16\epsilon$ independently of torsion, and the numerical integration with a running $\epsilon$ confirms that this conclusion does not depend on the constant-$\epsilon$ approximation. Torsion affects the ekpyrotic background more strongly than the spectra: it multiplies the steepness of a Jordan-frame exponential by $1+6\xi$ at large field and makes the Jordan-frame contraction cease to be ekpyrotic when $\sqrt\xi>(\lambda^2-6)/(4\lambda)$. In the entropic version the field space is asymptotically hyperbolic with curvature $-\xi$ when torsion is present, and the tilt of the entropy mode is shifted at order $\sqrt\xi$ by the non-minimal coupling and at order $\xi^{3/2}$ by torsion, which is below $10^{-3}$ for $\xi\le10^{-2}$. The regime where torsion matters in every sector examined is therefore the strong-field regime $\xi\phi^2\gtrsim1$.

The conclusions apply to the states, regimes and approximations that have been analysed. The wave functions are not normalisable on the minisuperspace, no physical inner product or boundary condition at $a=0$ is imposed, and ensembles are defined by the flux through a slice, so that the statements about bounces refer to the sample of 41 flux-weighted trajectories for one family of packets. The Laplace--Beltrami ordering is essential for the frame covariance, and a curvature term would break it. The perturbation theory is classical, with tensor modes transferred through the Bohmian bounce but scalar modes, the conversion of entropy perturbations and the amplitude and non-Gaussianity of the entropic spectrum left for a separate treatment, and the model is spatially flat, homogeneous and isotropic with at most two scalar fields. The observational comparison uses the Planck 2018 and BICEP/Keck 2021 results, and later data that favour a somewhat larger spectral index would shift the allowed window of $\xi$ without changing the comparison between torsion and the metric theory.

\appendix

\section{Riemann--Cartan decomposition for the homogeneous sector}
\label{app:RC}

For the metric \eqref{eq:flrw} write the connection as $\Gamma^\lambda{}_{\mu\nu}=\{^\lambda{}_{\mu\nu}\}+K^\lambda{}_{\mu\nu}$ with the contortion
\begin{equation}
K_{\lambda\mu\nu}=-\tfrac13\,(g_{\lambda\mu}T_\nu-g_{\nu\mu}T_\lambda)+\tfrac1{12}\,\epsilon_{\lambda\mu\nu\rho}S^\rho ,
\end{equation}
which is antisymmetric in $\lambda$ and $\nu$, so that the connection is metric compatible. With the conventions of Sec.~\ref{sec:formalism} the torsion is
$T^\lambda{}_{\mu\nu}=K^\lambda{}_{\mu\nu}-K^\lambda{}_{\nu\mu}$, its trace is $T^\nu{}_{\mu\nu}=T_\mu$ and $\epsilon^{\sigma\lambda\mu\nu}T_{\lambda\mu\nu}=S^\sigma$. For $T_\mu=(u,0,0,0)$ and $S_\mu=(v,0,0,0)$ the
Ricci scalar was computed symbolically for general $N(t)$ and $a(t)$ and gives
\begin{equation}
R(\Gamma)=\tilde R+\frac{2}{N^2}\Big[\frac{u^2}{3}-\dot u-\frac{3\dot a}{a}u+\frac{\dot N}{N}u\Big]-\frac{v^2}{24N^2},
\end{equation}
which equals $\tilde R+2\tilde\nabla_\mu T^\mu-\tfrac23T_\mu T^\mu+\tfrac1{24}S_\mu S^\mu$ with $\tilde\nabla_\mu T^\mu=-N^{-2}(\dot u+3\dot au/a-\dot Nu/N)$, $T_\mu T^\mu=-u^2/N^2$ and $S_\mu S^\mu=-v^2/N^2$. This
is the decomposition \eqref{eq:decomp}. With the coefficient $2/3$ and $F$ multiplying $R(\Gamma)$, the case $\kappa=0$ gives $K=1/F$ in \eqref{eq:K}, which is the known equivalence of Einstein--Cartan and
Palatini treatments of a non-minimally coupled scalar \cite{Shaposhnikov2020,Langvik2021}. This agreement serves as a check of the coefficients.

\section{Numerical methods}
\label{app:num}

Trajectories were obtained by integrating \eqref{eq:guid} with the explicit Runge--Kutta method of order 5(4) (scipy \texttt{solve\_ivp}), relative tolerance $10^{-9}$ to $10^{-10}$, absolute tolerance $10^{-11}$ to $10^{-12}$, and
maximum step $0.05$ to $0.1$ in $\tau$. The phase derivatives $S_{,\tilde\alpha}$ and $S_{,\chi}$ were evaluated from the analytic derivatives of \eqref{eq:twopacket}. The ensemble of Sec.~\ref{sec:massless} consists of
41 initial positions on $\tilde\alpha=3$ spaced uniformly over $\chi\in[-\sqrt6\cdot3\pm2.5\sqrt{2d}]$ with weights $\mathcal R^2|S_{,\tilde\alpha}|$; a trajectory was counted as a bounce if $\tilde\alpha$ has an interior minimum before
reaching $\tilde\alpha=-6$. The quantum term $\mathsf q$ was obtained by second differences of $|\Psi|$ with step $10^{-3}$, and the residual of the Hamilton--Jacobi identity was below the differencing error. The maps $\phi(\chi)$ for
$\kappa\neq0$ were tabulated on a logarithmic grid in $\phi$ by trapezoidal quadrature, and for $\kappa=0$ the closed form \eqref{eq:EC-map} was used. The mode equation of Sec.~\ref{sec:ekpyrosis} was integrated with an eighth-order Runge--Kutta method (relative tolerance $10^{-9}$) on a background interpolated by cubic splines in the number of e-folds, and the tensor equation \eqref{eq:tensorbounce} with the same method (relative tolerance $10^{-10}$). The Bessel-function solutions of Sec.~\ref{sec:exp} were checked with arbitrary-precision arithmetic (mpmath, 30 digits) by inserting them into \eqref{eq:wdwexp} with numerical derivatives, and the scaling attractor of Sec.~\ref{sec:exp} by integrating the Einstein-frame equations with relative tolerance $10^{-11}$. The slow-roll integrals of Sec.~\ref{sec:pert} were evaluated with
adaptive quadrature to relative accuracy $10^{-10}$, with $\phi_{\rm end}$ and $\phi_*$ found by root bracketing. The symbolic checks (Appendix~\ref{app:RC}, Eq.~\eqref{eq:Leinstein}, Eq.~\eqref{eq:RG}
and the conformal covariance of the Laplace--Beltrami operator) used sympy.

\end{document}